\documentclass[a4paper,UKenglish,cleveref, autoref, thm-restate]{lipics-v2021}

\usepackage{booktabs}

\usepackage{longtable}

\usepackage{todonotes}

\usepackage{epsfig}

\usepackage{proof}
\usepackage{mathpartir}
\usepackage{caption}
\usepackage{hyperref}

\usepackage{stmaryrd}

\usepackage{xcolor}

\usepackage{listings, listings-rust}
\usepackage{multicol}
\usepackage{array}
\usepackage{float}
\usepackage{wrapfig}

\usepackage{upquote,textcomp}

\DeclareSymbolFont{rsfs}{U}{rsfs}{m}{n}
\DeclareSymbolFontAlphabet{\mathscrsfs}{rsfs}

\definecolor{codegreen}{rgb}{0,0.6,0}
\definecolor{codegray}{rgb}{0.5,0.5,0.5}
\definecolor{codepurple}{rgb}{0.58,0,0.82}
\definecolor{codeblack}{rgb}{0,0,0}
\definecolor{backcolour}{rgb}{1,1,1}

\lstdefinestyle{basestyle}{
  backgroundcolor={},
  commentstyle=\color{codegreen},
  keywordstyle=\color{codeblack},
  numberstyle=\scriptsize\color{codeblack},
  stringstyle=\color{codepurple},
  breakatwhitespace=false,
  breaklines=true,
  captionpos=b,
  keepspaces=true,
  showspaces=false,
  showstringspaces=false,
  showtabs=false,
  tabsize=2,
  columns=fullflexible
}

\lstdefinestyle{mystyle}{
  backgroundcolor={},
  commentstyle=\color{codegreen},
  keywordstyle=\color{codeblack},
  numberstyle=\scriptsize\color{codeblack},
  stringstyle=\color{codepurple},
  basicstyle=\scriptsize\ttfamily,
  breakatwhitespace=false,
  breaklines=true,
  captionpos=b,
  keepspaces=true,
  showspaces=false,
  showstringspaces=false,
  showtabs=false,
  tabsize=2,
  columns=fullflexible
}

\lstdefinestyle{small-font}{
  basicstyle=\scriptsize\ttfamily,
  style=basestyle
}

\lstdefinestyle{normal-font}{
  basicstyle=\footnotesize\ttfamily,
  style=basestyle
}

\lstdefinestyle{normal-font-numbered}{
  numbers=left,
  numbersep=5pt,
  xleftmargin=5pt,
  xrightmargin=5pt,
  style=normal-font
}

\lstdefinestyle{small-font-numbered}
{ basicstyle=\scriptsize\ttfamily,%
  keywordstyle=\bfseries\color{PurpleKeyword},%
  keywordstyle=[3]\color{RedTypename},%
    comment=[l][\color{GrayComment}\slshape]{//},
    morecomment=[s][\color{GrayComment}\slshape]{/*}{*/},
    morecomment=[l][\color{GoldDocumentation}\slshape]{///},
    morecomment=[s][\color{GoldDocumentation}\slshape]{/*!}{*/},
    morecomment=[l][\color{GoldDocumentation}\slshape]{//!},
    morecomment=[s][\color{RedTypename}]{\#![}{]},
    morecomment=[s][\color{RedTypename}]{\#[}{]},
    stringstyle=\color{GreenString},
    string=[b]",
      numbers=left,
  numbersep=5pt,
  xleftmargin=10pt,
  xrightmargin=10pt
}

\lstdefinestyle{single-col}{
  style=normal-font
}

\lstdefinestyle{single-col-line-num}{
  backgroundcolor=\color{backcolour},
  commentstyle=\color{codegreen},
  keywordstyle=\color{codeblack},
  numberstyle=\scriptsize\color{codeblack},
  stringstyle=\color{codepurple},
  basicstyle=\footnotesize\ttfamily,
  breakatwhitespace=false,
  breaklines=true,
  captionpos=b,
  keepspaces=true,
  numbers=left,
  showspaces=false,
  showstringspaces=false,
  showtabs=false,
  tabsize=2,
  columns=fullflexible,
  escapechar=?
}

\newcommand{\code}[1]{\lstinline[language=rust, style=nicerust]{#1}}
\newcommand{\codee}[1]{{\small\texttt{#1}}}

\newcommand{\sendval}{\,\triangleleft\,}
\newcommand{\recvval}{\,\triangleright\,}

\newcommand{\lineartoshared}{\up^\mS_\mL}
\newcommand{\sharedtolinear}{\down^\mS_\mL}

\newcommand{\sills}{\m{SILL}_{\m{S}}}
\newcommand{\sillr}{\m{SILL}_{\m{R}}}
\newcommand{\SILLR}{$\m{SILL}_{\m{R}}\,$}

\usepackage{url}

\newcommand{\eg}{e.g.~}
\newcommand{\ie}{i.e.~}

\newcommand{\tdla}[1]{(\text{\textsc{T-#1$_{\mathrm{L}}$}})}

\newcommand{\tdra}[1]{(\text{\textsc{T-#1$_{\mathrm{R}}$}})}

\newcommand{\mi}[1]{\mathit{#1}}
\newcommand{\m}[1]{\mathsf{#1}}

\newcommand{\mb}[1]{\mathbf{#1}}
\newcommand{\semi}{\mathrel{;}}

\newcommand{\mmode}[1]{{\mathchoice{\m{#1}}{\m{#1}}{\scriptscriptstyle\m{#1}}{\scriptscriptstyle\m{#1}}}}
\newcommand{\mL}{\mmode{L}}
\newcommand{\mS}{\mmode{S}}

\newcommand{\D}{\Delta}

\newcommand{\entails}{\,\vdash\,}

\newcommand{\extchoice}{\binampersand}
\newcommand{\intchoice}{\oplus}

\newcommand{\chanout}{\otimes}
\newcommand{\chanin}{\multimap}
\newcommand{\up}{{\uparrow}}
\newcommand{\down}{{\downarrow}}

\newcommand{\pseq}[2]{#1 \, ; #2}

\newcommand{\pfwd}[2]{\m{fwd}\, #1 \, #2}

\newcommand{\pcutl}[3]{#1 \leftarrow #2 \leftarrow #3}

\newcommand{\pacq}[2]{#1 \leftarrow \m{acquire}\, #2}
\newcommand{\pacc}[2]{#1 \leftarrow \m{accept}\, #2}
\newcommand{\prel}[2]{#1 \leftarrow \m{release}\, #2}
\newcommand{\pdet}[2]{#1 \leftarrow \m{detach}\, #2}

\nolinenumbers

 \AtBeginDocument{%
   \providecommand\BibTeX{{%
     \normalfont B\kern-0.5em{\scshape i\kern-0.25em b}\kern-0.8em\TeX}}}

\title{Ferrite: A Judgmental Embedding of Session Types in Rust}

\author{Ruo Fei {Chen}}{Independent Researcher}{soares.chen@maybevoid.com}{https://orcid.org/0000-0001-5796-4386}{}
\author{Stephanie Balzer}{Carnegie Mellon University}{balzers@cs.cmu.edu}{}{National Science Foundation Award No. CCF-1718267}
\author{Bernardo Toninho}{Universidade Nova de Lisboa and NOVA
  LINCS}{btoninho@fct.unl.pt}{ https://orcid.org/0000-0002-0746-7514 }{FCT/MCTES grant NOVALINCS/BASE UIDB/04516/2020}

\authorrunning{R. F. Chen, S. Balzer, and B. Toninho}

\Copyright{R. F. Chen, S. Balzer, and B. Toninho}

\EventEditors{Karim Ali and Jan Vitek}
\EventNoEds{2}
\EventLongTitle{36th European Conference on Object-Oriented Programming (ECOOP 2022)}
\EventShortTitle{ECOOP 2022}
\EventAcronym{ECOOP}
\EventYear{2022}
\EventDate{June 6--10, 2022}
\EventLocation{Berlin, Germany}
\EventLogo{}
\SeriesVolume{222}
\ArticleNo{22}

\hypersetup{
pdftitle={Ferrite: A Judgmental Embedding of Session Types in Rust},
pdfsubject={cs.PL, cs.SE},
pdfauthor={Ruofei~Chen, Stephanie~Balzer, and Bernardo Toninho},
pdfkeywords={Session Types, Rust, DSL},
}



\begin{document}

\maketitle

\begin{abstract}
\emph{Session types} have proved viable in expressing and verifying
the protocols of message-passing systems. While message passing is a
dominant concurrency paradigm in practice, real world software is
written without session types.  A limitation of existing session type
libraries in mainstream languages is their restriction to linear
session types, precluding application scenarios that demand sharing
and thus aliasing of channel references.

This paper introduces Ferrite, a shallow embedding of session types in
Rust that supports both \emph{linear} and \emph{shared} sessions.  The
formal foundation of Ferrite constitutes the shared session type
calculus $\sills$, which Ferrite encodes via a novel \emph{judgmental
  embedding} technique.  The fulcrum of the embedding is the notion of
a typing judgment that allows reasoning about shared and linear
resources to type a session.  Typing rules are then encoded as
functions over judgments, with a valid typing derivation manifesting
as a well-typed Rust program.  This Rust program generated by Ferrite
serves as a \emph{certificate}, ensuring that the application will
proceed according to the protocol defined by the session type.  The
paper details the features and implementation of Ferrite and includes
a case study on implementing Servo's canvas component in Ferrite.

\end{abstract}

\section{Introduction}\label{sec:introduction}

Message-passing is a dominant concurrency paradigm, adopted by
mainstream languages such as Erlang, Scala, Go, and Rust, putting the
slogan \emph{``Do not communicate by sharing memory; instead, share
  memory by communicating''}~\cite{GoBlog2010} into practice.  In this
setting, messages are exchanged along channels, which can be shared by
several senders and receivers.  Type systems in such languages
typically allow channels to be typed, specifying and constraining the
types of messages they may carry (\eg integers, strings, sums,
references, etc.).

An aspect inherent to message-passing concurrency that is not captured
in mainstream type systems, however, is the idea of a \emph{protocol}.
Protocols dictate the sequencing and types of messages to be
exchanged.  To express and enforce such protocols, \emph{session
  types}~\cite{HondaCONCUR1993,HondaESOP1998,HondaPOPL2008} were
introduced.
Session typing disciplines assign types to channel endpoints according
to their intended usage protocols in terms of sequencing of
input/output actions (e.g.~``send an integer and, afterwards, receive
a string'') and branching/selection actions (e.g.~``receive either a
buy message and process the payment; or a cancellation message and
abort the transaction''), ensuring the action sequence is followed
correctly and thus, adherence to the protocol.
Thanks to their correspondence to \emph{linear}
logic~\cite{CairesCONCUR2010,WadlerICFP2012,ToninhoESOP2013,ToninhoPhD2015,LindleyESOP2015,CairesMSCS2016}
session types enjoy a strong logical foundation and ensure, in
addition to protocol adherence (\emph{session fidelity}), the
existence of a communication partner (\emph{progress}).
Session types have also been extended with safe
\emph{sharing}~\cite{BalzerICFP2017,BalzerCONCUR2018,BalzerESOP2019}
to accommodate multi-client scenarios that are rejected by exclusively
linear session types.

Despite these theoretical advances, session types have not (yet) been
adopted at scale.  While various session type embeddings exist in
mainstream languages such as Java~\cite{HuECOOP2008,HuECOOP2010},
Scala~\cite{ScalasECOOP2016},
Haskell~\cite{SackmanTR2008,PucellaHaskell2008,ImaiPLACES2010,LindleyHaskell2016},
OCaml~\cite{Padovani17,ImaiARTICLE2019}, and
Rust~\cite{JespersenWGP2015,KokkeICE2019,Rumpsteak2021Coord,Rumpsteak2021Arxiv}, all of these embeddings
lack support for multi-client scenarios that mandate controlled
aliasing in addition to linearity.
This paper introduces \emph{Ferrite}, a shallow embedding of session
types in Rust.  In contrast to prior work, Ferrite supports \emph{both}
linear and shared session types, with protocol adherence guaranteed
statically by the Rust compiler.

Ferrite's underlying theory is based on the calculus $\sills$ introduced in~\cite{BalzerICFP2017}, which develops the logical foundation of shared session types.
As a matter of fact, Ferrite encodes $\sills$ typing derivations
as Rust functions, through a technique we dub \emph{judgmental
  embedding}.  Through our judgmental embedding, a type-checked
Ferrite program yields a Rust program that corresponds to a $\sills$ typing
derivation and thus the \emph{proof} of protocol adherence.

In order to faithfully encode $\sills$ typing in Rust, this paper further makes
several technical contributions to emulate advanced typing
features, such as higher-kinded types, by a skillful combination of
traits (type classes) and associated types (type families).  For
example, Ferrite supports recursive (session) types in this way, which
are limited to recursive structs of a fixed size in plain Rust.
A combination of type-level natural numbers with ideas from profunctor
optics~\cite{OpticsPickeringGW17} are also used to support named
channels and labeled choices.  We adopt the idea of
\emph{lenses}~\cite{LensFosterGMPS07} for selecting and updating
individual channels in an arbitrary-length linear context.  Similarly,
we use \emph{prisms} for selecting a branch out of arbitrary-length
choices.  Whereas \code{session-ocaml} \cite{Padovani17} has previously explored the use of
n-ary choice through extensible variants in OCaml, we are the first to
connect n-ary choice to prisms and non-native implementation of
extensible variants.  Remarkably, the Ferrite codebase remains
entirely in the safe fragment of Rust, with no direct use of unsafe
features.

Given its support of both linear and shared session types, Ferrite is
capable of expressing any session typed program in Rust.
We
substantiate this claim by providing an implementation of Servo's
production canvas component with the communication layer entirely
within Ferrite.  We report on our findings, including benchmarks
in \Cref{sec:evaluation}.

In summary, this paper makes the following contributions:

\begin{itemize}

\item design and implementation of \emph{Ferrite}, an embedded domain-specific language (EDSL) for writing session-typed programs in Rust;

\item support of both \emph{linear} and \emph{shared} sessions, guaranteed to be observed by type checking;

 \item a novel \emph{judgmental embedding} of custom typing rules in a
 host language with the resulting program carrying the proof of
 successful type checking;

 \item an encoding of \emph{arbitrary-length choice} in terms of prisms and
 extensible variants in Rust;

\item an \emph{empirical evaluation} based on a full implementation of Servo's
 canvas component in Ferrite.

\end{itemize}

 \emph{Outline.}
\Cref{sec:background} gives a brief account of session types and
sharing, as found in the $\sills$ calculus~\cite{BalzerICFP2017}.
\Cref{sec:key-ideas} tours through the key ideas underlying Ferrite,
which are refined in subsequent sections.  \Cref{sec:statics}
introduces the technical aspects of Ferrite's type system, focusing on
the judgmental embedding and enforcement of linearity.
\Cref{sec:advanced} explains how Ferrite addresses Rust's limited
support of recursive data types to allow for arbitrary
recursive and shared session types.  \Cref{sec:choices} describes
the implementation of n-ary choice using prisms and extensible
variants.
\Cref{sec:evaluation} provides an evaluation of Ferrite via a re-implementation
of the Servo canvas component.
\Cref{sec:related} reports on related and future work.

An anonymized version of Ferrite's source code with examples is provided as an artifact.
All typing rules and their encoding as well as further materials of interest to an inquisitive reader are provided in the appendix.


\section{Background}\label{sec:background}

This section gives a brief tour of linear and shared session types.
The presentation is based on the intuitionistic session-typed process calculus $\sills$~\cite{BalzerICFP2017}, which Ferrite builds upon.
We consider the protocol governing the interaction between a queue and its client:

\begin{center}
\begin{small}
\begin{minipage}[t]{\textwidth}
\begin{tabbing}
$\m{queue}\; A$ $=$ \= $\extchoice \{$\= $\m{enq} : A \chanin \m{queue}\; A,$
$\m{deq} : \intchoice \{\m{none} : \mb{1}, \; \m{some} : A \chanout \m{queue}\; A\}\}$
\end{tabbing}
\end{minipage}
\end{small}
\end{center}

\Cref{tab:sills} provides an overview of the types used in the
example.  Since $\sills$ is based on a Curry-Howard correspondence
between intuitionistic linear logic and the session-typed
$\pi$-calculus~\cite{CairesCONCUR2010,CairesMSCS2016} it uses linear
logic connectives ($\intchoice$, $\extchoice$, $\chanout$, $\chanin$,
$\mb{1}$) as session types.  The remaining connectives concern shared
sessions, a feature we remark on shortly.
A crucial---and probably unusual---characteristic of session-typed
processes is that a process \emph{changes} its typing along with the
messages it exchanges.  As a result, a process' typing always reflects
the current protocol state.  \Cref{tab:sills} lists state transitions
inflicted by a message exchange in the first and second column and
corresponding process terms in the third and fourth column.  The fifth
column provides the operational meaning of a type.

\begin{table}
\caption{Overview of session types and terms in $\sills$ together with their operational meaning.
Subscripts $\m{L}$ and $\m{S}$ denote linear and shared sessions, resp., where $m, n \in \{ \m{L}, \m{S} \}$.}
\label{tab:sills}
\begin{small}
\begin{tabular}{@{}lllll@{}}
\toprule
\multicolumn{2}{@{}l}{\textbf{Session type}} & \multicolumn{2}{l}{\textbf{Process term}} & \\
\textbf{current} & \textbf{cont} & \textbf{current} &
\textbf{cont} & \textbf{Description} \\
\midrule
$c_\mL {:} \intchoice\{\overline{l {:} A_\mL}\}$ & $c_\mL {:} A_{\mL_{\mi{h}}}$ & $c_\mL.l_{\mi{h}} ; P$ &
$P$ & provider sends label $l_{\mi{h}}$ along $c_\mL$ \\
 & & $\m{case}\; c_\mL\; \m{of}\; \overline{l \Rightarrow Q}$ & $Q_{\mi{h}}$ & client receives label $l_{\mi{h}}$ along $c_\mL$ \\[2pt]
$c_\mL {:} \extchoice\{\overline{l {:} A_\mL}\}$ & $c_\mL {:} A_{\mL_{\mi{h}}}$ & $\m{case}\; c_\mL\; \m{of}\; \overline{l \Rightarrow P}$ &
$P_{\mi{h}}$ & provider receives label $l_{\mi{h}}$ along $c$ \\
 & & $c_\mL.l_{\mi{h}} ; Q$ & $Q$ & client sends label $l_{\mi{h}}$ along $c_\mL$ \\[2pt]
$c_\mL {:} A_m \chanout B_\mL$ & $c_\mL {:} B_\mL$ & $\m{send}\; c_\mL\; d_m ; P$ &
$P$ & provider sends channel $d_m {:} A_m$ along $c_\mL$ \\
 & & $y_m \leftarrow \m{recv}\; c_\mL ; Q_{\mi{y_m}}$ & $Q_{d_m}$ & client receives channel $d_m {:} A_m$ along $c_\mL$ \\[2pt]
$c_\mL {:} A_m \chanin B_\mL$ & $c_\mL {:} B_\mL$ & $y_m \leftarrow \m{recv}\; c_\mL ; P_{\mi{y_m}}$ &
$P_{d_m}$ & provider receives channel $d_m {:} A_m$ along $c_\mL$ \\
 & & $\m{send}\; c_\mL\; d_m ; Q$ & $Q$ & client sends channel $d_m {:} A_m$ along $c_\mL$ \\[2pt]
$c_\mL {:} \mb{1}$ & - & $\m{close}\; c_\mL$ &
- & provider sends ``$\m{end}$'' along $c_\mL$ \\
 & & $\m{wait}\; c_\mL ; Q$ & $Q$ & provider receives ``$\m{end}$'' along $c_\mL$ \\
$c_\mL {:} \down^\mS_\mL A_\mS$ & $c_\mS {:} A_\mS$ & $c_\mS \leftarrow \m{detach} \; c_\mL ; P_{\mi{x_\mS}}$ & 
$P_{c_\mS}$ & provider sends ``$\m{detach}\, c_\mS$'' along $c_\mL$ \\
 & & $x_\mS \leftarrow \m{release} \; c_\mL ; Q_{\mi{x_\mS}}$ & $Q_{c_\mS}$ & client receives ``$\m{detach}\, c_\mS$'' along $c_\mL$ \\
$c_\mS {:} \up^\mS_\mL A_\mL$ & $c_\mL {:} A_\mL$ & $c_\mL \leftarrow \m{acquire} \; c_\mS ; Q_{\mi{x_\mL}}$ & 
$Q_{c_\mL}$ & client sends ``$\m{acquire}\, c_\mL$'' along $c_\mS$ \\
 & & $x_\mL \leftarrow \m{accept} \; c_\mS ; P_{\mi{x_\mL}}$ & $P_{c_\mL}$ & provider receives ``$\m{acquire}\, c_\mL$'' along $c_\mS$ \\
$c_m : A_m$ & $c_m : A_m$ & $\pseq{\pcutl{z_n}{X}{\overline{d_m}}}{P_{z_n}}$ & $P_{z_n}$ & spawn ("cut") $X$ along $z_n {:} B_n$ with $\overline{d_m}{:}\overline{D_m}$ \\[2pt]
$c_m : A_m$ & - & $\pfwd{c_m}{d_m}$ & - & forward to channel $d_m {:} A_m$ and terminate \\[2pt]
\bottomrule
\end{tabular}
\end{small}
\end{table}

Consulting \Cref{tab:sills}, we gather that the above polymorphic
session type $\m{queue}\; A$ imposes the following recursive protocol:
A client may either send the label $\m{enq}$ or $\m{deq}$ to the
queue, depending on whether the client wishes to enqueue or dequeue an
element of type $A$, resp.  In the former case, the client sends the
element to be enqueued, after which the queue recurs.  In the latter
case, the queue indicates to the client whether it is empty
($\m{none}$) or not ($\m{some}$), and proceeds by either terminating
or sending the dequeued element and recurring, resp.

A linear typing discipline is beneficial because it immediately
guarantees session fidelity---even in the presence of perpetual
protocol change---by ensuring that a channel connects exactly two
processes.  Unfortunately, linearity also rules out various practical
programming scenarios that demand sharing and thus aliasing of channel
references.  For example, the above linear session type
$\m{queue}\; A$ is limited to a \emph{single} client.  To support safe
sharing of stateful channel references while upholding session fidelity,
$\sills$ extends linear session types with shared session types
($\down^\mS_\mL A_\mS$, $\up^\mS_\mL A_\mL$).  These two connectives
mediate between shared and linear sessions by requiring that clients
of shared sessions interact in \emph{mutual exclusion} from each
other.  Concretely, a type $\up^\mS_\mL A_\mL$ mandates a client to
\emph{acquire} the process offering the shared session.  If the
request is successful, the client receives a linear channel to the
acquired process along which it must proceed as detailed by the
session type $A_\mL$.  A type $\down^\mS_\mL A_\mS$, on the other
hand, mandates a client to \emph{release} the linear process,
relinquishing ownership of the linear channel and only being left with
a shared channel alias to the now shared process at type $A_\mS$.

Using these connectives, we can turn the above linear queue into a
shared one, bracketing enqueue and dequeue operations within
acquire-release:

\begin{center}
\begin{small}
\begin{minipage}[t]{\textwidth}
\begin{tabbing}
$\m{squeue}\; A_\mS$ $=$ \= $\up^\mS_\mL \extchoice \{$\= $\m{enq} : A_\mS \chanin \down^\mS_\mL \m{squeue}\; A_\mS,$
$\m{deq} : \intchoice \{\m{none} : \down^\mS_\mL \m{squeue}\; A_\mS, \; \m{some} : A_\mS \chanout \down^\mS_\mL \m{squeue}\; A_\mS\}\}$
\end{tabbing}
\end{minipage}
\end{small}
\end{center}
    
\noindent In contrast to the linear queue, the above version recurs in
the $\m{none}$ branch and thus keeps the queue alive to serve the next
client.  For convenience, $\sills$ allows the connectives $\chanout$
and $\chanin$ to be used to transport both linear and shared channels
along a linear carrier channel.

To provide a flavor of session-typed programming in $\sills$, we
briefly comment on the below processes $\mi{empty}$ and $\mi{elem}$,
which implement the shared queue session type as a sequence of
$\mi{elem}$ processes, ended by an $\mi{empty}$ process.  A process
implementation consists of its signature (first two lines) and body
(after {\small$=$}).  The first line indicates the typing of channel
variables used by the process (left of {\small$\vdash$}) and the type
of the providing channel variable (right of {\small$\vdash$}).  The
second line binds the channel variables.  In $\sills$,
{\small$\leftarrow$} generally denotes variable bindings.  We leave it
to the reader to convince themselves, consulting \Cref{tab:sills},
that the code in the body of the two processes executes the protocol
defined by session type $\m{squeue}\; A_\mS$.

\begin{center}
\begin{small}
\begin{minipage}[t]{0.5\textwidth}
\begin{tabbing}
$\cdot \vdash \mi{empty} :: q: \m{squeue}\; A_\mS$ \\
$q \leftarrow \mi{empty} \leftarrow \cdot = $ \\
\quad \= $q' \leftarrow \m{accept}\; q \semi$ \\
\> \= $\m{case}\; q'\; \m{of}$ \\
\> $\mid \m{enq} \rightarrow$ \= $x \leftarrow \m{recv}\; q' \semi$ \\
\> \> $q \leftarrow \m{detach}\; q' \semi$ \\
\> \> $e \leftarrow \mi{empty} \semi q \leftarrow \mi{elem}\leftarrow x, e$ \\
\> $\mid \m{deq} \rightarrow$ \> $q'.\m{none} \semi$ \\
\> \> $q \leftarrow \m{detach}\; q' \semi$ \\
\> \> $q \leftarrow \mi{empty}$
\end{tabbing}
\end{minipage}
\quad \quad \quad \quad
\begin{minipage}[t]{0.5\textwidth}
\begin{tabbing}
$x : A_\mS, t : \m{squeue}\; A_\mS \vdash \mi{elem} :: q : \m{squeue}\; A_\mS$ \\
$q \leftarrow \mi{elem} \leftarrow x, t = $ \\
\quad \= $q' \leftarrow \m{accept}\; q \semi$ \\
\> \= $\m{case}\; q'\; \m{of}$ \\
\> $\mid \m{enq} \rightarrow$ \= $y \leftarrow \m{recv}\; q' \semi$ \\
\> \> $t' \leftarrow \m{acquire} \; t \semi$ \\
\> \> $t'.\m{enq} \semi \m{send}\; t'\; y \semi$ \\
\> \> $t \leftarrow \m{release} \; t' \semi q \leftarrow \m{detach}\; q' \semi$ \\
\> \> $q \leftarrow \mi{elem} \leftarrow x, t$ \\
\> $\mid \m{deq} \rightarrow$ \> $q'.\m{some} \semi \m{send}\; q'\; x \semi$ \\
\> \> $q \leftarrow \m{detach}\; q' \semi \m{fwd}\; q \; t$ \\
\end{tabbing}
\end{minipage}
\end{small}
\end{center}

Imposing acquire-release not only as a programming methodology but
also as a \emph{typing discipline} has the advantage of recovering
session fidelity for shared sessions.  To this end, shared session
types in $\sills$ must be \emph{strictly
  equi-synchronizing}~\cite{BalzerICFP2017,BalzerESOP2019}, imposing
the invariant that an acquired session is released to the type at
which previously acquired.  For example, the shared session type
$\m{squeue}\; A_\mS$ is strictly equi-synchronizing whereas the type
$\m{invalid} = \up^\mS_\mL \extchoice \{ \m{left} : \down^\mS_\mL
\up^\mS_\mL \intchoice \{ \m{yes}: \down^\mS_\mL \m{invalid}, \m{no}:
\mb{1} \}, \m{right}: \down^\mS_\mL \m{invalid} \}$ is not.

It is instructive to review the typing rules for acquire-release:

\begin{small}
\begin{mathpar}
\inferrule[{\footnotesize\tdla{$\up^\mS_\mL$}}]
{\Psi, x_\mS : \up^\mS_\mL A_\mL; \D, y_\mL : A_\mL \entails Q_{y_\mL} :: (z_\mL : C_\mL)}
{\Psi, x_\mS : \up^\mS_\mL A_\mL; \D \entails \pseq{\pacq{y_\mL}{x_\mS}}{Q_{y_\mL}} :: (z_\mL : C_\mL)}

\inferrule[{\footnotesize\tdra{$\up^\mS_\mL$}}]
{\Psi; \cdot \entails P_{y_\mL} :: (y_\mL : A_\mL)}
{\Psi \entails \pseq{\pacc{y_\mL}{x_\mS}}{P_{y_\mL}} :: (x_\mS : \up^\mS_\mL A_\mL)}
\end{mathpar}
\end{small}

\begin{small}
\begin{mathpar}
\inferrule[{\footnotesize\tdla{$\down^\mS_\mL$}}]
{\Psi, x_\mS : A_\mS; \D \entails Q_{x_\mS} :: (z_\mL : C_\mL)}
{\Psi; \D, y_\mL : \down^\mS_\mL A_\mS \entails \pseq{\prel{x_\mS}{y_\mL}}{Q_{x_\mS}} :: (z_\mL : C_\mL)}

\inferrule[{\footnotesize\tdra{$\down^\mS_\mL$}}]
{\Psi \entails P_{x_\mS} :: (x_\mS : A_\mS)}
{\Psi; \cdot \entails \pseq{\pdet{x_\mS}{y_\mL}}{P_{x_\mS}} :: (y_\mL : \down^\mS_\mL A_\mS)}
\end{mathpar}
\end{small}

Due to its foundation in intuitionistic linear logic, $\sills$' typing
rules are phrased using a \emph{sequent calculus}, leading to
\emph{left} and \emph{right} rules for each connective.  Left rules
describe the interaction from the point of view of the client, right
rules from the point of view of the provider.  The typing judgments
$\Psi; \D \entails P :: (x_\mL : A_\mL)$ and
$\Psi \entails P :: (x_\mS : A_\mS)$ read as "process $P$ offers a
session of type $A$ along channel $x$ using sessions offered along
channels in $\Psi$ (and $\D$)."  The typing contexts $\Psi$ and
$\Delta$ provide the typing of shared and linear channels, resp.
Whereas $\Psi$ is a structural context, $\Delta$ is a linear context,
forbidding channels to be dropped (weakened) or duplicated
(contracted). In contrast to linear processes, shared processes must
not use any linear channels, a requirement crucial for type safety.
The notions of acquire and release are naturally formulated from the
point of view of a client, so these terms appear in the left rules.
The right rules use the terms \emph{accept} and \emph{detach} with the
meaning that an accept accepts an acquire and a detach initiates a
release.  The rules are read bottom-up, where the premise denotes the
next action to be taken after the message exchange.


\section{Key Ideas}\label{sec:key-ideas}




This section introduces the key ideas underlying Ferrite.
Subsequent sections provide further details.

\subsection{$\sillr$ -- A stepping stone from $\sills$ to Ferrite}

In \Cref{sec:background}, we reviewed $\sills$ and its typing
judgment. Our goal with Ferrite is to faithfully and compositionally
encode $\sills$ typing derivations in Rust.
However, when viewed under the lens of a general purpose programming
language, most readers will find $\sills$ a prohibitively austere
formalism, lacking most facilities needed to write realistic programs
(e.g.~basic data types, pattern matching, etc.) and provided by a convenient and usable programming
language like Rust. From an ergonomics standpoint alone it would be unreasonably
prohibitive for our embedding to forbid the use of Rust features such as
functions, traits and enumerations, only for the sake of precisely
mirroring $\sills$.
Moreover, to realize such an embedding we must be able to account for both
$\sills$' linear session discipline (i.e.~the \emph{linear} context
$\Delta$) and shared session discipline (i.e.~the \emph{structural} context $\Psi$) within Rust's usage discipline.
Since Rust's typing discipline is essentially \emph{affine}, its treatment of
variable usage is neither linear nor purely structural, and so both
shared and linear channels must be treated explicitly in the encoding.

The two points above naturally lead us to the language $\sillr$ as a
formal stepping stone between $\sills$ and our embedding, Ferrite.
$\sillr$ is, in its essence, a pragmatic extension of $\sills$ with
Rust (type and term) constructs, allowing us to intersperse
Rust code with the communication primitives of $\sills$.
In $\sillr$ we use the judgment
\begin{center}
\begin{small}
$\Gamma; \Delta \entails \mi{expr} :: A$,
\end{small}
\end{center}
denoting that expression $\mi{expr}$ has session type $A$, using the
sessions tracked by $\Gamma$ and $\Delta$.
This judgment differs from that of $\sills$
in its context region $\Gamma$ and term $\mi{expr}$, with the latter permitting arbitrary Rust expressions in addition to $\sills$ communication primitives. Whereas $\sills$'s
structural context $\Psi$ exclusively tracks shared channels,
$\sillr$'s $\Gamma$ tracks \emph{both} shared sessions (subject to
weakening and contraction) and plain Rust (affine) variables. A shared
channel type in both $\sillr$ and $\sills$ is always of the form
$\lineartoshared A$, so there is no confusion among the affine and
shared contents of $\Gamma$.
As we discuss
in  \Cref{sec:shared-session}, the distinction between a plain Rust
variable, which is treated as affine, and a shared channel, which is
treated structurally, is modelled in Ferrite by making
shared channels implement Rust's \code{Clone} trait.

\Cref{tab:ferrite} provides an overview of $\sillr$ types and terms and their Ferrite encoding.
$\sillr$ types stand in direct correspondence with $\sills$ types (see \Cref{tab:sills}), apart from shared channel output and input.
The $\sills$ types for sending and receiving shared channels
($A_\mS \chanout A_\mL$ and $A_\mS \chanin A_\mL$) correspond
to $\sillr$ types for sending and receiving values
($T \triangleleft A$ and $T \triangleright A$, resp.), which
support \emph{both} Rust values and shared channels.
Their typing rules are:

\vspace{-12pt}
\begin{small}
  \begin{mathpar}
\inferrule[(\textsc{T$\triangleleft_\m{R}$})]
{\Gamma \, ; \, \Delta \entails K :: A}
{\Gamma, \, x: \tau ; \, \Delta \entails
  \m{send\_value} \; x ; \, K :: \tau \sendval A}
\qquad
\inferrule[(\textsc{T$\triangleleft_\m{L}$})]
{ \Gamma, \, \mi{x}: \tau \, ; \Delta, a: A \entails
  K \, :: \, B
}
{ \Gamma \, ; \Delta, a: \tau \recvval A \entails
  \mi{x} \leftarrow \m{receive\_value\_from} \, a; \, K \, :: \, B
}
\end{mathpar}
\vspace{-8pt}
\end{small}

\noindent
Rule \textsc{T$\triangleleft_\m{R}$} indicates that the
value bound to variable $x$ of type $\tau$ will be sent, after which
the continuation $K$ will execute, offering type $A$. Dually, rule
\textsc{T$\triangleleft_\m{L}$} states that using such a provider
bound to $a$ will bind $x$ of type $\tau$ in continuation $K$, which
must now use the channel bound to $a$ according to $A$.

\begin{table}
\caption{Overview of $\sillr$ types and terms and their encoding in Ferrite.
Note that $\sillr$ uses $\mi{\tau} \sendval A_\mL$ and $\mi{\tau} \recvval A_\mL$ for shared channel output and input, resp., and $\epsilon$ for termination.}
\label{tab:ferrite}
\begin{small}
\begin{tabular}{@{}llll@{}}
\toprule
\multicolumn{2}{@{}l}{\textbf{Type}} & \multicolumn{2}{l}{\textbf{Terms ($\sillr$)}} \\
\textbf{Ferrite} & \textbf{$\sillr$} & \textbf{provider} & \textbf{client} \\
\midrule
\code{InternalChoice<Row>} & $\intchoice\{ \overline{ l_i: A_{\mL_i} } \}$ & $\m{offer} \; l_i; K$ & $\m{case} \; a \; \{ \overline{ l_i: K_i } \}$ \\
\code{ExternalChoice<Row>} & $\extchoice\{ \overline{ l_i: A_{\mL_i} } \}$ & $\m{offer\_choice} \{ \overline{ l_i: K_i } \}$ & $\m{choose} \; a \; l_i ; K$ \\
\code{SendChannel<A,B>} & $A_\mL \chanout B_\mL$ & $\m{send\_channel\_from} \; a; K$ & $a \leftarrow \m{receive\_channel\_from} \; f \; a; K$ \\
\code{ReceiveChannel<A,B>} & $A_\mL \chanin B_\mL$ & $a \leftarrow \m{receive\_channel}; K$ & $\m{send\_channel\_to} \; f \; a; K$ \\
\code{SendValue<T,A>} & $\mi{\tau} \sendval A_\mL$ & $\m{send\_value} \; x; K$ & $x \leftarrow \m{receive\_value\_from} \, a \, x; K$ \\
\code{ReceiveValue<T,A>} & $\mi{\tau} \recvval A_\mL$ & $x \leftarrow \m{receive\_value}; K$ & $\m{send\_value\_to} \; a \; x; K$ \\
\code{End} & $\mb{\epsilon}$ & $\m{terminate}$ & $\m{wait} \; a; K$ \\
\code{SharedToLinear<A>} & $\sharedtolinear A_\mS$ & $\m{detach\_shared\_session}; K_s$ & $\m{release\_shared\_session} \, a; K_l$ \\
\code{LinearToShared<A>} & $\lineartoshared A_\mL$ & $\m{accept\_shared\_session}; K_l$ & $a \leftarrow \m{acquire\_shared\_session} \, s; K_l$ \\
\bottomrule
\end{tabular}
\end{small}
\end{table}

\subsection{Judgmental Embedding}\label{sec:key-ideas-judgmental-embedding}

Having introduced the $\sillr$ typing judgment and illustrated some
of its typing rules, we can now clarify the idea behind our notion of
\emph{judgmental embedding}, which enables the Rust compiler to
typecheck $\sillr$ programs by encoding typing derivations as Rust
programs.  The basic idea underlying this encoding can be
schematically described as follows:

\begin{minipage}[c]{0.5\textwidth}

\begin{center}
\begin{small}
$
\infer
{\Gamma \, ; \Delta_1 \entails \mi{expr}; \mi{cont} :: A_1}
{\Gamma \, ; \Delta_2 \entails \mi{cont} :: A_2}
$
\end{small}
\end{center}

\end{minipage}
\begin{minipage}[c]{0.5\textwidth}

\begin{lstlisting}[language=Rust, style=nicerust]
fn expr<...>
  ( cont: PartialSession<C2, A2> )
  -> PartialSession<C1, A1>
\end{lstlisting}

\end{minipage}

\noindent On the left we show a $\sillr$ typing rule and on the right
its encoding in Ferrite.
Ferrite encodes a $\sillr$ \emph{typing judgment} $\Gamma; \Delta
\entails \mi{expr} :: A$ as a value of Rust \emph{type}
\code{PartialSession<C, A>}, where \code{C} encodes the linear context
$\Delta$ and \code{A} the session type $A$, standing for any of the Ferrite types of \Cref{tab:ferrite}.
Ferrite then encodes a $\sillr$ \emph{typing rule} for an expression $\mi{expr}$ as a Rust \emph{function} \code{expr}
that accepts a \code{PartialSession<C2, A2>} and returns a
\code{PartialSession<C1, A1>}, where $\mi{expr}$ stands for any of the $\sillr$ terms of \Cref{tab:ferrite}.
The encoding makes use of
\emph{continuation passing style} (arising from the sequent
calculus-based formulation of $\sillr$), with the return type
being the conclusion of the rule and the argument type
being its premise.
\Cref{tab:sillr-embedding} summarizes the judgmental embedding;
\Cref{sec:statics-judgment} provides further details.  Whereas Ferrite
explicitly performs a type-level encoding of the linear context
$\Delta$, the representation of the shared and affine context region
$\Gamma$ is achieved through Rust's normal binding structure, with the
obligation that shared channels implement Rust's \code{Clone} trait to
permit contraction.  To type a closed program, Ferrite defines the
type \code{Session<A>}, which stands for a $\sillr$ judgment with an
empty linear context.

\begin{table}[t]
\caption{Judgmental embedding of $\sillr$ in Ferrite.}
\label{tab:sillr-embedding}
\centering
\begin{footnotesize}
\begin{tabular}{@{}lll@{}}
\toprule
\textbf{$\sillr$} & \textbf{Ferrite} &  \textbf{Description} \\[3pt]
\hline
$\Gamma \, ; \, \cdot \, \vdash \, A$ & \texttt{Session<A>}
& Typing judgment for top-level session (\ie closed program). \\
$\Gamma \, ; \, \Delta \, \vdash \, A$ & \texttt{PartialSession<C, A>}
& Typing judgment for partial session. \\
$\Delta$ & \texttt{C: Context} & Linear context; explicitly encoded.  \\
$\Gamma$ & - & Shared / Affine context; delegated to Rust. \\
$A$ & \texttt{A: Protocol} & Session type. \\
\bottomrule
\end{tabular}
\end{footnotesize}
\vspace{-10pt}
\end{table}

Adopting a judgmental embedding technique for implementing a DSL delivers the benefits of proof-carrying code:
the \code{PartialSession<C1, A1>} returned from a well-typed Ferrite \code{expr} \emph{is} the typing derivation of the corresponding $\sillr$ term.
In case the $\sillr$ term is a $\sills$ term, its typing derivation certifies protocol adherence by virtue of the type safety proof of $\sills$~\cite{BalzerICFP2017}.
In case the $\sillr$ term includes Rust code, its typing derivation certifies protocol adherence modulo the possibility of a panic raised by the Rust code.

\subsection{Recursive and Shared Session Types in Ferrite}

Rust's support for recursive types is limited to recursive struct
definitions of a known size.  To circumvent this restriction and
support arbitrary recursive session types, Ferrite introduces a
type-level fixed-point combinator \code{Rec<F>} to obtain the fixed
point of a type function \code{F}.  Since Rust lacks higher-kinded
types such as \code{Type} $\rightarrow$ \code{Type}, we use
\emph{defunctionalization}~\cite{ReynoldsACM1972,HKTYallopW14} by
accepting any Rust type \code{F} implementing the trait \code{RecApp}
with a given associated type \code{F::Applied}, as shown below.
 \Cref{sec:recursive-types} provides further details.

\begin{lstlisting}[language=Rust, style=nicerust]
trait RecApp<X> { type Applied; }
struct Rec<F: RecApp<Rec<F>>> { unfold: Box<F::Applied> }
\end{lstlisting}

Recursive types are also vital for encoding shared session types.  In
line with~\cite{BalzerESOP2019}, we restrict shared session
types to be recursive, making sure that a shared component is
continuously available.  To guarantee type preservation, recursive
session types must be \emph{strictly
  equi-synchronizing}~\cite{BalzerICFP2017,BalzerESOP2019}, requiring
an acquired session to be released to the same type at which it was
previously acquired.  Ferrite enforces this invariant by defining a
specialized trait \code{SharedRecApp} which omits an implementation
for \code{End}:

\begin{lstlisting}[language=Rust, style=nicerust]
trait SharedRecApp<X> { type Applied; }   trait SharedProtocol { ... }
struct SharedToLinear<F> { ... }    struct SharedChannel<S: SharedProtocol> { ... }
struct LinearToShared<F: SharedRecApp<SharedToLinear<LinearToShared<F>>>> { ... }
\end{lstlisting}

Ferrite achieves safe communication for shared sessions by imposing an
acquire-release discipline~\cite{BalzerICFP2017} on shared sessions,
establishing a critical section for the linear portion of the process
enclosed within the acquire and release.  \code{SharedChannel} denotes
the shared process running in the background, and clients with a
reference to it can \textit{acquire} an exclusive linear channel to
communicate with it.  As long as the linear channel exists, the shared
process is locked and cannot be acquired by any other client.  With
the strictly equi-synchronizing constraint in place, the now linear
process must eventually be released (\code{SharedToLinear}) back to
the same shared session type at which it was previously acquired,
giving turn to another client waiting to acquire.
 \Cref{sec:shared-session} provides further details on the encoding.

\subsection{N-ary Choice and Linear Context}

Ferrite implements n-ary \emph{choices} and linear typing
\emph{contexts} as extensible \emph{sums} and \emph{products} of
session types, resp.  Ferrite uses heterogeneous
lists~\cite{HList} to annotate a list of session types of arbitrary
length.  The notation $\m{HList!}[A_0, A_1, ..., A_{N-1}]$ denotes a
heterogeneous list of N session types, with $A_i$ being the session
type at the $i$-th position of the list. The $\m{HList!}$ macro acts
as syntactic sugar for the heterogeneous list, which in its raw form
is encoded as $(A_0, (A_1, (..., (A_{N-1}, ()))))$. Ferrite uses the
Rust tuple constructor \code{(,)} for $\m{HCons}$, and unit \code{()}
for $\m{HNil}$.  The heterogeneous list itself can be directly used to
represent an n-ary product.  Using an associated type, the list can
moreover be transformed into an n-ary sum.

One disadvantage of using heterogeneous lists is that its elements
have to be addressed by position rather than a programmer-chosen
label. To recover labels for accessing list elements, we use
optics~\cite{OpticsPickeringGW17}.  More precisely, Ferrite uses
\emph{lenses}~\cite{LensFosterGMPS07} to access a channel in a linear
context and \emph{prisms} to select a branch of a choice.  We further
combine the optics abstraction with \emph{de Bruijn levels} and
implement lenses and prisms using type level natural numbers. Given an
inductive trait definition of natural numbers as zero (\code{Z}) and
successor (\code{S<N>}), a natural number \code{N} implements the lens
to access the N-th element in the linear context, and the prism to
access the N-th branch in a choice.  Schematically, the lens encoding
can be captured as follows:

\begin{minipage}[c]{0.4\textwidth}

\begin{center}
\begin{small}
$
\inferrule[]
{\Gamma \, ; \Delta, \, l_n : B_2\entails \mi{K} :: A_2}
{\Gamma \, ; \Delta, \, l_n : B_1 \entails \mi{expr} \; l_n; \mi{K} :: A_1}
$
\end{small}
\end{center}

\end{minipage}
\begin{minipage}[c]{0.6\textwidth}

\begin{lstlisting}[language=Rust, style=nicerust]
fn expr<...>
  ( l: N, cont: PartialSession<C1, A2> )
  -> PartialSession<C2, A1>
where N: ContextLens<C1, B1, B2, Target=C2>
\end{lstlisting}

\end{minipage}

\noindent The index \code{N} amounts to the type of the variable
\code{l} that the programmer chooses as a name for a channel in the
linear context.  Ferrite handles the mapping, supporting random
access to programmer-named channels.
 \Cref{sec:statics-linear-context} provides further details,
including the support of higher-order channels.  Similarly, prisms
allow choice selection in constructs such as \code{offer\_case} to be
encoded as follows:

\begin{minipage}[c]{0.45\textwidth}

\begin{center}
\begin{small}
$
\infer
{\Gamma ; \Delta \entails \m{offer\_case} \; l_n; \; \mi{K} :: \intchoice\{..., \, l_n: A_n, ... \}}
{\Gamma ; \Delta  \entails \mi{K} :: A_n}
$
\end{small}
\end{center}

\end{minipage}
\begin{minipage}[c]{0.55\textwidth}

\begin{lstlisting}[language=Rust, style=nicerust]
fn offer_case<N, Row, C, A>
  ( l: N, cont: PartialSession<C, A> )
  -> PartialSession<C, InternalChoice<Row>>
where N: Prism<Row, Elem=A>, ...
\end{lstlisting}

\end{minipage}

Ferrite maps a choice label to a constant having the singleton value
of a natural number \code{N}, which implements the prism to access the
N-th branch of a choice.  In addition to prisms, Ferrite implements a
version of \textit{extensible variants}~\cite{Morris15} to support
polymorphic operations on arbitrary sums of session types representing
choices. Finally, the \code{define\_choice!} macro is used as a helper
to export type aliases as programmer-friendly identifiers.
Details 
are reported in \Cref{sec:choices}.


\section{Ferrite -- A Judgmental Embedding of $\sillr$}\label{sec:statics}

Having introduced some of the key concepts to the implementation of
Ferrite, we now cover in detail the implementation of Ferrite's core
constructs, building up the knowledge required for
\Cref{sec:advanced} and \Cref{sec:choices}.
Ferrite, like any other DSL, has to tackle the various technical
challenges encountered when embedding a DSL in a host language.  In
doing so, we take inspiration from the range of embedding techniques
developed for Haskell and adjust them to the Rust setting. The lack of
higher-kinded types, limited support of recursive types, and presence
of weakening, in particular, make the development far from trivial.  A
more conceptual contribution of this work is thus to demonstrate how
existing Rust features can be combined to emulate many of the missing
features that are beneficial to DSL embeddings and how to encode
custom typing rules in Rust or any similarly expressive language.
The techniques described in this and subsequent sections also serve as
a reference for embedding other DSLs in a host language like Rust.

\subsection{Encoding Typing Rules via Judgmental Embedding}\label{sec:statics-judgment}

A distinguishing characteristic of Ferrite is its \emph{propositions
  as types} approach, yielding a direct correspondence between
$\sillr$ notions and their Ferrite encoding.  This correspondence was
introduced in \Cref{sec:key-ideas-judgmental-embedding} (see
\Cref{tab:sillr-embedding}) and we now discuss it in more detail.
To this end, let's consider the typing of value input.  We remind the
reader of \Cref{tab:ferrite} in \Cref{sec:key-ideas},
which provides a mapping between $\sillr$ and Ferrite session types.
Interested readers can find a corresponding mapping on the term level
in \Cref{tab:term-mapping} in the
supplement. 

\begin{center}
\begin{small}
$
\infer[(\textsc{T$\recvval_\m{R}$})]
{ \Gamma \, ; \Delta \entails
  \mi{a} \leftarrow \m{receive\_value}; \, K \, :: \,
  \tau \recvval A
}
{
  \Gamma, \, \mi{a}: \tau \, ; \Delta \entails
  K \, :: \, A
}
$
\end{small}
\end{center}

\noindent The $\sillr$ right rule \textsc{T$\recvval_\m{R}$} types the
expression $\mi{a} \leftarrow \m{receive\_value}; \, K $ as the
session type $\tau \recvval A$ and the continuation $K$ as the session
type $A$, where $a$ is now in scope with type $\tau$.
Following the schema hinted in
\Cref{sec:key-ideas-judgmental-embedding}, Ferrite encodes this rule
as the function \code{receive\_value}, parameterized by a value type
\code{T} ($\tau$), a linear context \code{C} ($\Delta$), and an
offered session type \code{A}.

\begin{lstlisting}[language=Rust, style=nicerust]
fn receive_value<T, C:Context, A:Protocol>(cont:impl FnOnce(T) -> PartialSession<C, A>)
                                   -> PartialSession<C, ReceiveValue<T, A>>
\end{lstlisting}

The function yields a value of type \code{PartialSession<C,
  ReceiveValue<T, A>>}, i.e. the conclusion of the rule, given an
(affine) closure of type \code{T} $\rightarrow$
\code{PartialSession<C, A>}, encoding the premise of the
rule. Notably, Ferrite uses plain Rust binding (through function
types) to encode the contents of $\Gamma$, as illustrated for the
received value above. The use of a closure reveals the
continuation-passing-style of the encoding, where the received value
of type \code{T} is passed to the continuation closure.  The affine
closure implements the \code{FnOnce} trait, ensuring that it can only
be called once.

The type \code{PartialSession} is a core construct of Ferrite that
enables the judgmental embedding of $\sillr$.  A Rust value of type
\code{PartialSession<C, A>} represents a Ferrite program that
guarantees linear usage of session type channels in the linear context
\code{C} and offers the linear session type \code{A}, corresponding to
the $\sillr$ typing judgment $\Gamma ; \Delta \vdash \mi{expr} :: A$.
The type parameters \code{C} and \code{A} are constrained to implement
the traits \code{Context} and \code{Protocol} -- two other Ferrite
constructs representing a linear context and linear session type,
resp.:

\begin{lstlisting}[language=Rust, style=nicerust]
trait Context { ... }     trait Protocol { ... }
struct PartialSession<C: Context, A: Protocol> { ... }
\end{lstlisting}

For each $\sillr$ session type, Ferrite defines a corresponding Rust
struct that implements the trait \code{Protocol}, yielding the listing
shown in \Cref{tab:ferrite}.  Implementations for
$\epsilon$ (\code{End}) and $\tau \recvval A$ (\code{ReceiveValue<T,
  A>}) are shown below.  When a session type is nested within another
session type, such as in the case of \code{ReceiveValue<T, A>}, the
constraint to implement \code{Protocol} is propagated to the inner
session type, requiring \code{A} to also implement \code{Protocol}:

\begin{lstlisting}[language=Rust, style=nicerust]
struct End { ... }    struct ReceiveValue<T, A> { ... }
impl Protocol for End { ... }
impl<A: Protocol> Protocol for ReceiveValue<T, A> { ... }
\end{lstlisting}

Thus, while Ferrite delegates the handling of the shared/structural
context $\Gamma$ to Rust, the encoding of the linear context $\Delta$
is explicit.  Being affine, the Rust type system permits weakening, a
structural property rejected by linear logic. Ferrite encodes a
linear context as a heterogeneous (type-level) list~\cite{HList} of
the form \codee{HList![A$_0$, A$_1$, ..., A$_{N-1}$]}, with all its
type elements \codee{A$_i$} implementing \code{Protocol}.
Internally, the \code{HList} macro desugars the type-level list into a
nested tuple \codee{(A$_0$, (A$_1$, (..., (A$_{N-1}$, ()))))}.  The
unit type \code{()} is used as the empty list (\code{HNil}) and the
tuple constructor \code{(,)} is used as the \code{HCons} constructor.
The implementation for \code{Context} is defined inductively as
follows:

\begin{lstlisting}[language=Rust, style=nicerust]
impl Context for () { ... }   impl<A: Protocol, C: Context> Context for (A, C) { ... }
\end{lstlisting}

To represent a closed program, i.e. a program without free channel variables,
we define a type alias \code{Session<A>} for
\code{PartialSession<C, A>}, with \code{C} restricted to the empty
context:

\begin{lstlisting}[language=Rust, style=nicerust]
type Session<A> = PartialSession<(), A>;
\end{lstlisting}

A complete session type program in Ferrite is thus of
type \code{Session<A>} and amounts to the $\sillr$ typing derivation
proving that the program adheres to the defined protocol.  Below we
show a ``hello world''-style program in Ferrite:

\begin{lstlisting}[language=Rust, style=nicerust]
let hello_provider = receive_value(|name| {
  println!("Hello, {}", name); terminate() });
\end{lstlisting}

The Ferrite program \code{hello\_provider} has an inferred Rust type
\code{Session<ReceiveValue<String, End>>}. It offers the
type \code{ReceiveValue<String, End>} by first receiving a string
value using \code{receive\_value}, binding it to
\code{name} in the continuation closure. Upon receiving the name
string, It prints out the name with a \code{"Hello"} greeting, and
terminates using \code{terminate()}.

\subsection{Manipulating the Linear Context}\label{sec:statics-linear-context}

\paragraph*{Context Lenses}
The use of a type-level list to encode the linear context has the
advantage of allowing contexts of arbitrary length. However, the list
imposes an order on the context's elements, disallowing exchange.  To
allow exchange, we make use of the concept of
\textit{lenses}~\cite{LensFosterGMPS07} to define a \code{ContextLens}
trait, which is implemented using type-level natural numbers.

\begin{lstlisting}[language=Rust, style=nicerust]
#[derive(Copy)] struct Z;     #[derive(Copy)] struct S<N>(PhantomData<N>);
trait ContextLens<C: Context, A1: Protocol, A2: Protocol> { type Target: Context; ... }
\end{lstlisting}

The \code{ContextLens} trait defines the read and update operations on a linear context,
such that given a \textit{source} context \codee{C = HList![..., A$_N$, ...]},
the source element of interest, \codee{A$_N$} at position $N$, can be updated to
the target element \code{B}
to form the \textit{target} context \code{Target =  HList![..., B, ...]},
with the remaining elements unchanged.
We use natural numbers to inductively implement \code{ContextLens}
at each position in the linear context, such that it satisfies all constraints
of the form:
\begin{center}
\codee{N: ContextLens<HList![..., A$_N$, ...], A$_N$, B, Target=HList![..., B, ...]>}
\end{center}
The implementation of natural numbers as context lenses is done by
first considering the base case, with \code{Z} used to access
the first element of any non-empty linear context:

\begin{lstlisting}[language=Rust, style=nicerust]
impl<A1: Protocol, A2: Protocol, C: Context> ContextLens<(A1, C), A1, A2>
  for Z { type Target = ( A2, C ); ... }
impl<A1: Protocol, A2: Protocol, B: Protocol, C: Context, N: ContextLens<C, A1, A2>>
 ContextLens <(B, C), A1, A2> for S<N> { type Target = (B, N::Target); ... }
\end{lstlisting}

In the inductive case, for any natural number \code{N} implementing
the context lens for a context \codee{HList![A$_0$, ..., A$_{N}$,
  ...]}, it's successor \code{S<Z>} implements the context lens
for \codee{HList![A$_{-1}$, A$_0$, ..., A$_{N}$, ...]}, with a new element
\codee{A$_{-1}$} appended to the head of the linear context.  Using
context lenses, we can encode the $\sillr$ left rule
$\textsc{T$\triangleright_\m{L}$}$ shown below, which types sending an
ambient value $x$ to a channel $a$ in the linear context that expects
to receive a value.

\begin{center}
{\small
$
\inferrule*[right=(\textsc{T$\triangleright_\m{L}$})]
{\Gamma \, ; \, \Delta, a: A \entails K :: B}
{\Gamma, \, x: \tau ; \, \Delta, \, a: \tau \recvval A \entails
  \m{send\_value\_to} \; a \; x ; \, K :: B}
$}
\end{center}
In Ferrite, $\textsc{T$\triangleright_\m{L}$}$ is implemented as the function
\code{send\_value\_to}, which uses a context lens \code{N} to send a value
of type \code{T} to the \code{N}-th channel in the linear context \code{C1}.
This requires the \code{N}-th channel to have type \code{ReceiveValue<T,A>}.
A continuation \code{cont} is then given with the linear context \code{C2},
which has the \code{N}-th channel updated to type \code{A}.

\begin{lstlisting}[language=Rust, style=nicerust]
fn send_value_to<N, T, C1: Context, C2: Context, A: Protocol, B: Protocol>
  ( n: N, x: T, cont: PartialSession<C2, B> ) -> PartialSession <C1, B>
where N: ContextLens<C1, ReceiveValue<T, A>, A, Target=C2>
\end{lstlisting}

\paragraph*{Channel Removal}

The above definition of a context lens is suited for \emph{updating}
channel types in a context. However, we have not addressed how
channels can be \emph{removed} or \emph{added} to the linear context.
These operations are required to implement session termination and
higher-order channel constructs such as $\chanout$ and $\chanin$.  To
support channel removal, we introduce a special \code{Empty} element
to denote the \textit{absence} of a channel at a given position in the
linear context:

\begin{lstlisting}[language=Rust, style=nicerust]
struct Empty;     trait Slot { ... }
impl Slot for Empty { ... }     impl<A: Protocol> Slot for A { ... }
\end{lstlisting}

To allow \code{Empty} to be present in a linear context,
we introduce a new \code{Slot} trait and make both
\code{Empty} and \code{Protocol} implement \code{Slot}.
The original definition of \code{Context} is then updated
to allow types that implement \code{Slot} instead of
\code{Protocol}.
\begin{center}
{\small
$
\inferrule*[right=(\text{\textsc{T1$_\m{L}$}})]
{ \Gamma \, ; \, \Delta  \entails K :: A
}
{ \Gamma \, ; \, \Delta, \, a: \epsilon \entails \m{wait} \, a; \, K :: A
}
\qquad
\inferrule*[right=(\text{\textsc{T1$_\m{R}$}})]
{ }
{ \Gamma \, ; \cdot \entails
  \m{terminate}; \, :: \, \epsilon
}
$}
\end{center}
Using \code{Empty}, it is straightforward to implement $\sillr$'s
session termination. Rule \textsc{T1$_\m{L}$} is encoded via a context
lens that replaces a channel of session type \code{End} with the
\code{Empty} slot. The function \code{wait} shown below does not
really remove a slot from a linear context, but merely replaces the
slot with \code{Empty}.  The use of \code{Empty} is necessary, because
we want to preserve the position of channels in a linear context in
order for the context lens for a channel to work across continuations.

\begin{lstlisting}[language=Rust, style=nicerust]
fn wait<C1: Context, C2: Context, A: Protocol, N>
  ( n: N, cont: PartialSession<C2, A> ) -> PartialSession<C1, A>
where N: ContextLens<C1, End, Empty, Target=C2>
\end{lstlisting}

With \code{Empty} introduced, an empty linear context may now contain
any number of \code{Empty} slots, such as \code{HList![Empty, Empty]}.
We introduce a new \code{EmptyContext} trait to abstract over the
different forms of empty linear contexts and provide an inductive
definition as its implementation:

\begin{lstlisting}[language=Rust, style=nicerust]
trait EmptyContext: Context { ... }   impl EmptyContext for () { ... }
impl<C: EmptyContext> EmptyContext for (Empty, C) { ... }
\end{lstlisting}

Given the empty list \code{()} as the base case, the
inductive case \code{(Empty, C)} is an empty linear context, if
\code{C} is also an empty linear context.
Using the definition of an empty context, the $\sillr$ right rule
$\textsc{T}\textbf{1}_\m{R}$ can then be easily encoded as the function
\code{terminate}, which works generically for all contexts
that implement \code{EmptyContext} as shown below:

\begin{lstlisting}[language=Rust, style=nicerust]
fn terminate<C: EmptyContext>() -> PartialSession<C, End>
\end{lstlisting}

\paragraph*{Channel Addition}

The Ferrite function \code{wait} removes a channel from the linear
context by replacing it with \code{Empty}. Dually, the function
\code{receive\_channel}, adds a new channel to the
linear context.  The \SILLR rule \textsc{T$\chanin_\m{R}$} for
channel input is shown below.  It binds the received channel of
session type $A$ to the channel variable $a$ and adds it to the linear
context $\Delta$ of the continuation.

\begin{center}
 {\small
$
\inferrule*[right=($\textsc{T$\chanin_\m{R}$}$)]
{\Gamma \, ; \, \Delta, a: A \entails K :: B}
{ \Gamma \, ; \, \Delta \entails
  a \, \leftarrow \, \m{receive\_channel} ; \, K :: A \chanin B
}
$}
\end{center}

To encode \textsc{T$\chanin_\m{R}$}, an append operation on contexts is defined
via the \code{AppendContext} trait:

\begin{lstlisting}[language=Rust, style=nicerust]
trait AppendContext<C: Context>: Context { type Appended: Context; ... }
impl<C: Context> AppendContext<C> for () { type Appended = C; ... }
impl<A: Slot, C1: Context, C2: Context, C3: Context> AppendContext<C2>
  for (A, C1) where C1: AppendContext<C2, Appended=C3> { type Appended = (A, C3); ... }
\end{lstlisting}

The \code{AppendContext} trait is parameterized by a linear context
\code{C} and an associated type \code{Appended}.  If a linear context
\code{C1} implements the trait \code{AppendContext<C2>}, it means that
context \code{C2} can be appended to \code{C1}, with \code{C3} =
\code{C1::Appended} being the result of the append operation.  The
implementation of \code{AppendContext} is defined inductively, with
the empty list \code{()} implementing the base case and the cons cell
\code{(A, C)} implementing the inductive case.

Using \code{AppendContext}, a channel \code{B} can be appended to the
end of a linear context \code{C}, if \code{C} implements
\code{AppendContext<HList![B]>}. The new linear context after the
append operation is given in the associated type
\code{C::Appended}.  We then observe that the position of channel
\code{B} in \code{C::Appended} is the same as the length of the
original linear context \code{C}. In other words, the context lens for
channel \code{B} in \code{C::Appended} can be generated by obtaining
the length of \code{C}.
In Ferrite, the length operation is implemented by
adding an associated type \code{Length} to the \code{Context}
trait. The implementation of \code{Context} for
\code{()} and \code{(A, C)} is updated correspondingly.

\begin{lstlisting}[language=Rust, style=nicerust]
trait Context { type Length; ... }    impl Context for () { type Length = Z; ... }
impl<A: Slot, C: Context> Context for (A, C) { type Length = S<C::Length>; ... }
\end{lstlisting}

The $\sillr$ right rule \textsc{T${\chanin_\m{R}}$}
is then encoded  as follows:

\begin{lstlisting}[language=Rust, style=nicerust]
fn receive_channel<A: Protocol, B: Protocol, C1: Context, C2: Context>(
  cont: impl FnOnce(C1::Length) -> PartialSession<C2, B>) ->
  PartialSession<C1, ReceiveChannel<A, B>> where C1: AppendContext<(A, ()), Appended=C2>
\end{lstlisting}

The function \code{receive\_channel} is parameterized by a linear
context \code{C1} implementing \code{AppendContext} to append the
session type \code{A} to \code{C1}. The continuation argument
\code{cont} is a closure that is given a context lens
\code{C::Length}, and returns a \code{PartialSession} with
\code{C2=C1::Appended} as its linear context. The function returns a
\code{PartialSession} with linear context \code{C1}, offering session
type \code{ReceiveChannel<A, B>}.

It is worth noting that in the type signature of \code{receive\_channel},
the type \code{C1::Length} is not shown to have any \code{ContextLens} implementation.
However when \code{C1::Length} is instantiated into the concrete types
\code{Z}, \code{S<Z>}, etc in the continuation body, Rust will use the
appropriate implementations of \code{ContextLens} so that they can
be used to access the appended channel in the linear context.

The use of \code{receive\_channel} is illustrated with the
\code{hello\_client} example below:

\begin{lstlisting}[language=Rust, style=nicerust]
let hello_client = receive_channel(|a| {
  send_value_to(a, "Alice".to_string(), wait(a, terminate())) });
\end{lstlisting}

The \code{hello\_client} program is inferred to have the Rust type
\code{Session<ReceiveChannel< ReceiveValue<String, End>, End>>}.  It
is written to communicate with the \code{hello\_provider} program
defined earlier in \Cref{sec:statics-judgment}.
The interaction is achieved by having \code{hello\_client} offering
the session type \code{ReceiveChannel<ReceiveValue<String, End>,
  End>}.  In its body, \code{hello\_client} uses
\code{receive\_channel} to receive channel \code{a} of type
\code{ReceiveValue<String, End>} from \code{hello\_provider}. The
continuation closure is given an argument \code{a:Z}, denoting the
context lens generated by \code{receive\_channel} for accessing the
received channel in the linear context.  The context lens \code{a:Z}
is then used for sending a string value, after which we \code{wait}
for \code{hello\_provider} to terminate.  We note that the type
\code{Z} of channel \code{a} (\ie the channel position in the context)
is automatically inferred by Rust and not exposed to the user.

\subsection{Communication}\label{sec:communication}

At this point we have defined the necessary constructs to build and
typecheck both \code{hello\_provider} and \code{hello\_client}, but
the two are separate Ferrite programs that are yet to be linked with
each other and executed.

\begin{minipage}[c]{0.5\textwidth}
\begin{small}
\begin{center}
\[
\inferrule*[right=(\text{\textsc{T-cut}})]
{ \Gamma \, ; \, \Delta_1 \entails K_1 :: A
  \and
  \Gamma \, ; \, \Delta_2, a: A \entails K_2 :: B
}
{
  \Gamma \, ; \, \Delta_1, \Delta_2 \entails
  a \, \leftarrow \, \m{cut} \; K_1 \, ; \; K_2 \, :: \, B
}
\]
\end{center}
\end{small}

\end{minipage}
\begin{minipage}[c]{0.5\textwidth}
\begin{small}
\begin{center}
\[\qquad
\inferrule*[right=(\text{\textsc{T-fwd}})]
{ \, }
{ \Gamma \, ; a: A \entails
  \m{forward} \, a \, :: \, A
}
\]
\end{center}
\end{small}
\end{minipage}

\vspace{5pt}

In \SILLR, rule \textsc{T-cut} allows two session-typed programs to
run in parallel, with the channel offered by $K_1$ added to the linear
context of program $K_2$.
Together with the forward rule \textsc{T-fwd}, we can use cut twice to
run both \code{hello\_provider} and \code{hello\_client} in parallel,
and have a third program that sends the channel offered by
\code{hello\_provider} to \code{hello\_client}. The program
\code{hello\_main} would have the following pseudo code in \SILLR:

\vspace{3pt}
\begin{minipage}[c]{0.5\textwidth}
$
\begin{array}{ll}
\mi{hello\_main} : \, \epsilon \; = & \mi{f} \, \leftarrow \, \m{cut} \; \mi{hello\_client} ; \\

& \mi{a} \, \leftarrow \, \m{cut} \; \mi{hello\_provider} ; \\

& \m{send\_channel\_to} \; \mi{f} \; \mi{a} ; \\

& \m{forward} \; \mi{f}
\end{array}
$
\end{minipage}
\vspace{3pt}

To implement \code{cut} in Ferrite, we need a way to split a linear
context \code{C} = $\Delta_1,
\Delta_2$ into two sub-contexts \code{C1} =
$\Delta_1$ and \code{C2} = $\Delta_2$ so that they can be passed to
the respective continuations. Moreover, since Ferrite programs use
context lenses to access channels, the ordering of channels inside
\code{C1} and \code{C2} must be preserved. We can preserve the
ordering by replacing the corresponding slots with \code{Empty} during
the splitting. Ferrite defines the \code{SplitContext} trait to
implement the splitting as follows:

\begin{lstlisting}[language=Rust, style=nicerust]
enum L {}     enum R {}
trait SplitContext<C: Context> { type Left: Context; type Right: Context; ... }
\end{lstlisting}

We first define two (uninhabited) marker types \code{L} and \code{R}.
We then use type-level lists consisting of elements \code{L} and
\code{R} to implement the \code{SplitContext} trait for a given linear
context \code{C}. The \code{SplitContext} implementation contains the
associated types \code{Left} and \code{Right}, representing the
contexts \code{C1} and \code{C2} after splitting. As an example, the
type \code{HList![L, R, L]} would implement
\code{SplitContext<HList![A1, A2, A3]>} for any slot \code{A1},
\code{A2} and \code{A3}, with the associated type \code{Left} being
\code{HList![A1, Empty, A3]} and \code{Right} being
\code{HList![Empty, A2, Empty]}.  We omit the implementation details
of \code{SplitContext} for brevity. Using \code{SplitContext}, the
function \code{cut} can be implemented as follows:

\begin{lstlisting}[language=Rust, style=nicerust]
fn cut<XS, C: Context, C1: Context, C2: Context, C3: Context, A: Protocol, B: Protocol>
  ( cont1: PartialSession<C1, A>,
    cont2: impl FnOnce(C2::Length) -> PartialSession<C3, B> ) -> PartialSession<C, B>
where XS: SplitContext<C, Left=C1, Right=C2>, C2: AppendContext<HList![A], Appended=C3>
\end{lstlisting}

The function \code{cut} works by using the heterogeneous list
\code{XS} that implements \code{SplitContext} to split a linear
context \code{C} into \code{C1} and \code{C2}.  To pass on the channel
\code{A} that is offered by \code{cont1} to \code{cont2}, \code{cut}
uses a similar technique to \code{receive\_channel} to append the
channel \code{A} to the end of \code{C2}, resulting in
\code{C3}. Using \code{cut}, we can write \code{hello\_main} in
Ferrite as follows:

\begin{lstlisting}[language=Rust, style=nicerust]
let hello_main: Session<End> =
  cut::<HList![]>(hello_client, |f| { cut::<HList![R]>(hello_provider, |a| {
    send_channel_to(f, a, forward(f)) }) });
\end{lstlisting}

Due to ambiguous instances for \code{SplitContext}, the type parameter
\code{XS} has to be annotated explicitly for Rust to know in which
context a channel should be placed.  In the first use of \code{cut},
the context is empty, so we call \code{cut} with the empty list
\code{HList![]}. We pass \code{hello\_client} as the first
continuation to run in parallel, and name the channel offered by
\code{hello\_client} as \code{f}.  In the second use of \code{cut},
the linear context would be \code{HList![ReceiveValue<String, End>]},
with one channel \code{f}. We then have \code{cut} move \code{f} to
the right side using \code{HList![R]}. On the left continuation, we
have \code{hello\_provider} run in parallel, and name the offered
channel as \code{a}. In the right continuation, we use
\code{send\_channel\_to} to send channel \code{a} to
\code{f}. Finally, we forward the continuation of \code{f}, which now
has type \code{End}.

Although \code{cut} provides the primitive way for Ferrite programs to
communicate, its use can be cumbersome and requires a lot of
boilerplate. For simplicity, Ferrite provides a specialized construct
\code{apply\_channel} that abstracts over the common pattern usage of
\code{cut} described earlier.  \code{apply\_channel} takes a client
program \code{f} offering session type \code{ReceiveChannel<A, B>} and
a provider program \code{a} offering session type \code{A}, and sends
\code{a} to \code{f} using \code{cut}.  The use of
\code{apply\_channel} is akin to regular function application,
making it more intuitive for programmers to use:
\begin{lstlisting}[language=Rust, style=nicerust]
fn apply_channel<A: Protocol, B: Protocol>(
  f: Session<ReceiveChannel<A, B>>, a: Session<A>) -> Session<B>
\end{lstlisting}

\subsection{Executing Ferrite Programs}

To actually \emph{execute} a Ferrite program, the program must offer
some specific session types. In the simplest case, Ferrite provides
the function \code{run\_session} for running a top-level Ferrite
program offering \code{End}, with an empty linear context:
\begin{lstlisting}[language=Rust, style=nicerust]
async fn run_session(session: Session<End>) { ... }
\end{lstlisting}

Function \code{run\_session} executes the session
\textit{asynchronously} using Rust's async/await
infrastructure. Internally, the struct \code{PartialSession<C, A>}
implements the dynamic semantics of the Ferrite program, which is only
accessible by public functions such as \code{run\_session}. Ferrite
currently uses the \code{tokio}~\cite{TokioWebsite} runtime for
asynchronous execution, as well as the one shot channels from
\code{tokio::sync::oneshot} to implement the low-level communication
of Ferrite channels.

Since \code{run\_session} accepts an argument of type
\code{Session<End>}, this means that programmers must first use
\code{cut} or \code{apply\_channel} to fully link partial Ferrite
programs with free channel variables, or Ferrite programs that offer
session types other than \code{End} before they can be executed.  This
restriction ensures that all linear channels created by a Ferrite
program are consumed.  For example, the programs
\code{hello\_provider} and \code{hello\_client} cannot be executed
individually, but the linked program resulting from composing
\code{hello\_provider} with \code{hello\_client} can be executed:

\begin{lstlisting}[language=Rust, style=nicerust]
async fn main() { run_session(apply_channel(hello_client, hello_provider)).await; }
\end{lstlisting}

We omit the implementation details of the dynamics of
Ferrite, which use low-level primitives such as Rust channels while
carefully ensuring that the requirements and invariants of session
types are satisfied. Interested readers can find more details in \Cref{sec:dynamics}.


\section{Recursive and Shared Session Types}\label{sec:advanced}

Many real world applications, such as web services and instant
messaging, implement protocols that are recursive in nature.  As a
result, it is essential for Ferrite to support recursive session
types.  In this section, we first report on Rust's limited support for
recursive types and how Ferrite addresses this limitation.  We then
discuss how Ferrite encodes \emph{shared} session types, which are
recursive.

\subsection{Recursive Session Types}\label{sec:recursive-types}

Consider a simple example of a counter session type, which sends an
infinite stream of integer values, incrementing each by one. To write
a Ferrite program that offers such a session type, we may attempt to
define the counter session type as follows:

\begin{lstlisting}[language=Rust, style=nicerust]
type Counter = SendValue<u64, Counter>;
\end{lstlisting}

\noindent If we try to use the type definition above, the compiler
will emit the error "cycle detected when processing \code{Counter}".
The problem with the above definition is that it is a directly
self-referential type alias, which is not supported in Rust.  Rust
imposes various restrictions on the legal forms of recursive types to
ensure that the memory layout of data is known at compile-time.

\paragraph*{Type-Level Fixed Points}

To address this limitation, we implement type-level fixed points using
\emph{defunctionalization}~\cite{ReynoldsACM1972,HKTYallopW14}. 
This is done by introducing a \code{RecApp} trait that is implemented
by defunctionalized types that can be ``applied'' with a type
parameter:

\begin{lstlisting}[language=Rust, style=nicerust]
trait RecApp<X> { type Applied; }   type AppRec<F, X> = <F as RecApp<X>>::Applied;
struct Rec<F: RecApp<Rec<F>>> { unfold: Box<AppRec<F, Rec<F>>> }
\end{lstlisting}

The \code{RecApp} trait is parameterized by a type \code{X}, which
serves as the type argument to be applied to.  This makes it possible
for a Rust type \code{F} that implements \code{RecApp} to act as if it
has the higher-kinded type $\m{Type} \rightarrow \m{Type}$, and be
``applied'' to type \code{X}. We define a type alias \code{AppRec<F, X>} to
refer to the associated type \code{Applied} resulting from
``applying'' \code{F} to \code{X} via \code{RecApp}.  Using
\code{RecApp}, we can now define a type-level recursor \code{Rec} as a struct parameterized
by a type \code{F} that implements \code{RecApp<Rec<F>>}.  The body
of \code{Rec} contains a boxed value \code{Box<AppRec<F,
  RecApp<Rec<F>>>>} to make it have a fixed size in Rust.

Ferrite implements \code{RecApp} for all \code{Protocol} types, with the
type \code{Z} used to denote the recursion point. With that, the
example \code{Counter} type would be defined as follows:

\begin{lstlisting}[language=Rust, style=nicerust]
type Counter = Rec<SendValue<u64, Z>>;
\end{lstlisting}

The type \code{Rec<SendValue<T, Z>>} is unfolded into
\code{SendValue<T, Rec<SendValue<T, Z>>>}. This is achieved by
having the following generic implementations of \code{RecApp} for
\code{SendValue} and \code{Z}:

\begin{lstlisting}[language=Rust, style=nicerust]
impl<X> RecApp<X> for Z { type Applied = X; }
impl<X, T, A: RecApp<X>> RecApp <X> for SendValue <T, A> {
  type Applied = SendValue<T, AppRec<A, X>; }
\end{lstlisting}

Inside \code{RecApp}, \code{Z} simply replaces itself with the type argument
\code{X}.  \code{SendValue<T, A>} delegates the type
application of \code{X} to \code{A}, provided that the session type
\code{A} also implements \code{RecApp} for \code{X}.

The session type \code{Counter} is iso-recursive, as the rolled type
\code{Rec<SendValue<u64, Z>>} and the folded type
\code{SendValue<u64, Rec<SendValue<u64, Z>>} are considered distinct
types in Rust. As a result, Ferrite provides the constructs
\code{fix\_session} and \code{unfix\_session} for converting
between the rolled and unfolded versions of a recursive session type.

\paragraph*{Nested Recursive Session Types}

The use of \code{RecApp} is akin to emulating the higher-kinded type (HKT)
$\m{Type} \rightarrow \m{Type}$ in Rust. As of this writing, HKTs are
only available in the nightly (unstable) version of Rust through
\textit{generic associated types}.
However even with support for HKTs, our defunctionalization-based
approach via \code{RecApp} allows us to generalize to
\textit{nested} recursive types.

To account for a recursive type with multiple recursion points, we
introduce a \textit{recursion context} \code{R} as a type-level list
of elements (c.f. the linear context of
\Cref{sec:statics-linear-context}).  The type-level natural numbers
\code{Z}, \code{S<Z>}, etc. are now used as de Bruijn indices to unfold
to the elements in the recursion context.  The type-level fixed point
combinator \code{Rec} is redefined as \code{RecX}, containing the
recursion context:

\begin{lstlisting}[language=Rust, style=nicerust]
struct RecX<R, F: RecApp<(RecX<R, F>, R)>> { unfix: Box<AppRec<F, (RecX<R, F>, R)>> }
type Rec<F> = RecX<(), F>;
impl<R, F: RecApp<(RecX<R, F>, R)>> RecApp<R> for RecX<(), F>  {
    type Applied = RecX<R, F>; }
\end{lstlisting}

A recursive session type is defined starting with an empty recursion
context. Since nested recursive session types allow a \code{RecX} to be
embedded inside another \code{RecX}, we have \code{RecX} also implement
\code{RecApp}, provided it has an empty recursion context.  When
unfolded from another recursion context \code{R}, \code{RecX} simply
saves \code{R} as its own recursion context and does not unfold
further in \code{F}.  The inner type \code{F} is only unfolded once
with the full recursion context after all surrounding \code{RecX} types
are unfolded.

The recursive marker \code{Z} is modified to unfold to the first element
of the recursion context. We then implement \code{S<N>} to unfold to
the (N+1)-th position in the recursion context:

\begin{lstlisting}[language=Rust, style=nicerust]
impl<A, R> RecApp<(A, R)> for Z { type Applied = A; }
impl<A, R, N: RecApp<R>> RecApp<(A, R)> for S<N> { type Applied = N::Applied; }
\end{lstlisting}


\subsection{Shared Session Types}\label{sec:shared-session}

In the previous section we explored a recursive session type
\code{Counter}, which is defined using \code{Rec} and \code{Z}.  Since
\code{Counter} is defined as a linear session type, it cannot be
shared among multiple clients.  Shared communication, however, is
essential to implement many practical applications.  For instance, we
may want to implement a simple counter web-service, to send a unique
count for each request. To support such shared communication, we
introduce \textit{shared session types} in Ferrite, enabling
\emph{safe} shared communication in the presence multiple clients.

\paragraph*{Shared Session Types in Ferrite}

As introduced in \Cref{sec:background}, the $\sills$ (and $\sillr$)
notion of shared session types is recursive in nature, as a shared
session type must offer the same linear critical section to all
clients that acquire a shared resource. For instance, a shared version
of the \code{Counter} type in $\sillr$ is:
\begin{center}
\begin{small}
$
\m{SharedCounter} = \lineartoshared \m{Int} \sendval \sharedtolinear \m{SharedCounter}
$
\end{small}
\end{center}
 The linear
portion of $\m{SharedCounter}$ in between $\lineartoshared$ (acquire)
and $\sharedtolinear$ (release) amounts to a critical section.  When a
$\m{SharedCounter}$ is \textit{acquired}, it offers a linear session
type $\m{Int} \sendval \sharedtolinear \m{SharedCounter}$, willing to
send an integer value, after which it must be \emph{released} to
become available again as a $\m{SharedCounter}$ to the next client.

The recursive aspect of shared session types in $\sillr$ means that we
can reuse the implementation technique that we use for recursive
session types.  The type \code{SharedCounter} can be
defined in Ferrite as follows:
\begin{lstlisting}[language=Rust, style=nicerust]
type SharedCounter = LinearToShared<SendValue<u64, Release>>;
\end{lstlisting}

Compared to linear recursive session types, the main difference is
that instead of using \code{Rec}, a shared session type is defined
using the \code{LinearToShared} construct. This corresponds to
$\lineartoshared$ in \SILLR, with the inner type \code{SendValue<u64,
  Release>} corresponding to the linear portion of the shared session
type. At the point of recursion, the type \code{Release} is used in
place of $\sharedtolinear \m{SharedCounter}$.  As a result, the type
\code{LinearToShared<SendValue<u64, Release>>} is unfolded into
\code{SendValue<u64, SharedToLinear<LinearToShared<SendValue<u64,
  Release>>>>} after being acquired.  Type unfolding is implemented as
follows:

\begin{lstlisting}[language=Rust, style=nicerust]
trait SharedRecApp<X> { type Applied; }   trait SharedProtocol { ... }
struct SharedToLinear<F> { ... }          struct LinearToShared<F> { ... }
impl<F> Protocol for SharedToLinear<LinearToShared<F>>
  where F: SharedRecApp<SharedToLinear<LinearToShared<F>>> { ... }
impl<F> SharedProtocol for LinearToShared<F>
  where F: SharedRecApp<SharedToLinear<LinearToShared<F>>> { ... }
\end{lstlisting}

The struct \code{LinearToShared} is parameterized by a linear session
type \code{F} that implements the trait
\code{SharedRecApp<SharedToLinear<LinearToShared<F>>>}.  It uses the
\code{SharedRecApp} trait instead of the \code{RecApp} trait to ensure
that the session type is \textit{strictly
  equi-synchronizing}~\cite{BalzerESOP2019}, requiring an acquired
session to be released to the same type at which it was previously
acquired. Ferrite enforces this requirement by omitting an
implementation of \code{SharedRecApp} for \code{End}, ruling out
invalid shared session types such as
\code{LinearToShared<SendValue<u64, End>>}.  We note that the type
argument to \code{F}'s \code{SharedRecApp} is another struct
\code{SharedToLinear}, which corresponds to $\sharedtolinear$ in
\SILLR.  A \code{SharedProtocol} trait is also defined to
identify shared session types, \ie \code{LinearToShared}.

Once a shared process is started, a shared channel is created to allow
multiple clients to access the shared process through the use of shared
channel:

\begin{lstlisting}[language=Rust, style=nicerust]
struct SharedChannel<S: SharedProtocol> { ... }
impl<S> Clone for SharedChannel<S> { ... };
\end{lstlisting}

The code above shows the definition of the \code{SharedChannel}
struct. Unlike linear channels, shared channels follow structural
typing, \ie they can be weakened or contracted.  This means that we
can delegate the handling of shared channels to Rust, given that
\code{SharedChannel} implements Rust's \code{Clone} trait to allow
contraction.  Whereas $\sills$ provides explicit constructs for
sending and receiving shared channels, Ferrite's shared channels can
be sent as regular Rust values using \codee{Send/ReceiveValue}.

On the client side, a \code{SharedChannel} serves as an endpoint for
interacting with a shared process running in parallel. To start the
execution of such a shared process, a corresponding Ferrite program
has to be defined and executed. Similar to \code{PartialSession}, we
define \code{SharedSession} as shown below to represent such a shared
Ferrite program.

\begin{lstlisting}[language=Rust, style=nicerust]
struct SharedSession<S: SharedProtocol> { ... }
fn run_shared_session<S: SharedProtocol>(session: SharedSession<S>) -> SharedChannel<S>
\end{lstlisting}

Just as \code{PartialSession} encodes linear Ferrite programs without
executing them, \code{SharedSession} encodes shared Ferrite programs
without executing them.  Since \code{SharedSession} does not implement
the \code{Clone} trait, the shared Ferrite program is itself affine
and cannot be shared.  To enable sharing, the shared Ferrite program
must first be executed with \code{run\_shared\_session}.  The function
\code{run\_shared\_session} takes a shared Ferrite program of type
\code{SharedSession<S>} and starts it in the background as a shared
process. Then, in parallel, the shared channel of type
\code{SharedChannel<S>} is returned to the caller, which can then be
sent to multiple clients for access to the shared process.

The details of each shared Ferrite construct are
described in \Cref{sec:shared-ferrite-constructs}.  Below
we demonstrate how a program with a shared session can be defined and
used by multiple clients:

\begin{lstlisting}[language=Rust, style=nicerust]
type SharedCounter = LinearToShared<SendValue<u64, Release>>;
fn counter_producer(current_count: u64) -> SharedSession<SharedCounter> {
  accept_shared_session(async move {
    send_value(current_count, detach_shared_session(
        counter_producer(current_count + 1))) }) }

fn counter_client(counter: SharedChannel<SharedCounter>) -> Session<End> {
  acquire_shared_session(counter, move | chan | {
    receive_value_from(chan, move | count | {
      println!("received count: {}", count);
      release_shared_session(chan, terminate()) }) }) }
\end{lstlisting}

The recursive function \code{counter\_producer} creates a
\code{SharedSession} program that, when executed, offers a shared
channel of session type \code{SharedCounter}. On the provider side, a
shared session is defined using the \code{accept\_shared\_session}
construct, with a continuation given as an async thunk that is
executed when a client acquires the shared session and enters the
linear critical section (of type \code{SendValue<u64,
  SharedToLinear<SharedCounter>>}).
Inside the closure, the producer uses \code{send\_value} to send the
current count to the client and then uses
\code{detach\_shared\_session} to exit the linear critical section.
The construct \code{detach\_shared\_session} offers the linear session
type \code{SharedToLinear<SharedCounter>} and expects a continuation
that offers the shared session type \code{SharedCounter} to serve the
next client. We generate the continuation by recursively calling the
\code{counter\_producer} function.

The \code{counter\_client} function takes a shared channel of session
type \code{SharedCounter} and returns a session type program that
acquires the shared channel and prints the received count value to the
terminal. A linear Ferrite program can acquire a shared session using
the \code{acquire\_shared\_session} construct, which accepts a
\code{SharedChannel} object and adds the acquired linear channel to
the linear context.
In this case, the continuation closure is given the context lens
\code{Z}, which provides access to the linear channel of session type
\code{SendValue<u64, SharedToLinear<SharedCounter>>} in the first slot
of the linear context. It then uses \code{receive\_value\_from} to
receive the value sent by the shared provider and then prints the
value.  On the client side, the linear session of type
\code{SharedToLinear<SharedCounter>} must be released using the
\code{release\_shared\_session} construct. After releasing the shared
session, other clients will then be able to acquire the shared
session.

\begin{lstlisting}[language=Rust, style=nicerust]
async fn main () {
  let counter1: SharedChannel<SharedCounter> = run_shared_session(counter_producer(0));
  let counter2 = counter1.clone();
  let child1 = task::spawn(async move { run_session(counter_client(counter1)).await; });
  let child2 = task::spawn(async move { run_session(counter_client(counter2)).await; });
  join!(child1, child2).await; }
\end{lstlisting}

To illustrate a use of \code{SharedCounter}, we have a \code{main}
function that initializes a shared producer with an initial value of 0
and then runs the shared provider 
using the \code{run\_shared\_session} construct.  The returned
\code{SharedChannel} is then cloned, making the shared counter
accessible via aliases \code{counter1} and \code{counter2}.  It then
uses \code{task::spawn} to spawn two async tasks that run
\code{counter\_client} twice.  A key observation is that multiple
Ferrite programs that are executed independently can access \emph{the same}
shared producer through a reference to the shared channel.



\section{N-ary Choice}\label{sec:choices}

Session types support \emph{internal} and \emph{external} choice,
leaving the choice among several options to the provider or the
client, resp. (see \Cref{tab:ferrite}).  When
restricted to binary choice, the implementation is relatively
straightforward, as shown below by the two right rules for internal
choice in \SILLR.  The $\m{offer\_left}$ and $\m{offer\_right}$
constructs allow a provider to offer an internal choice
$A \intchoice B$ by offering either $A$ or $B$, resp.

\vspace{-8pt}
{\small
\[
\inferrule*[right=($\textsc{T$\intchoice\m{2}\m{L}_\m{R}$}$)]
{ \Gamma \, ; \, \Delta \entails K :: A
}
{ \Gamma \, ; \, \Delta \entails
  \m{offer\_left}; \, K :: A \intchoice B
}
\qquad
\inferrule*[right=($\textsc{T$\intchoice\m{2}\m{R}_\m{R}$}$)]
{ \Gamma \, ; \, \Delta \entails K :: B
}
{ \Gamma \, ; \, \Delta \entails
  \m{offer\_right}; \, K :: A \intchoice B
}
\]}
\vspace{-8pt}

It is straightforward to implement the two versions of the right rules
by writing the two respective functions \code{offer\_left} and \code{offer\_right}:

\begin{lstlisting}[language=Rust, style=nicerust]
fn offer_left<C: Context, A: Protocol, B: Protocol>
  ( cont: PartialSession<C, A> ) -> PartialSession<C, InternalChoice2<A, B>>
fn offer_right < C: Context, A: Protocol, B: Protocol >
  ( cont: PartialSession<C, B> ) -> PartialSession<C, InternalChoice2<A, B>>
\end{lstlisting}

However, this approach does not scale if we want to generalize choice
beyond two options. To support N-ary choice, the functions would
have to be explicitly reimplemented N times. Instead, we implement a
single \code{offer\_case} function which allows selection
from 
n-ary branches.

\subsection{Prisms}

In \Cref{sec:statics-linear-context}, we explored heterogeneous list
to encode the linear context, \ie \textit{products} of session types
of arbitrary lengths.  We then implemented context \emph{lenses} to
access and update individual channels in the linear context.
Observing that n-ary choices can be encoded as \textit{sums} of
session types, we now use \emph{prisms} to implement the selection of
an arbitrary-length branch.
Instead of having a binary choice type \code{InternalChoice2<A, B>},
we can define an n-ary choice type \code{InternalChoice<HList![...]>},
with \code{InternalChoice<HList![A, B]>} being the special case of a
binary choice. To select a branch out of the heterogeneous list, we
define the \code{Prism} trait as follows:

\begin{lstlisting}[language=Rust, style=nicerust]
trait Prism<Row> { type Elem; ... }
impl<A, R> Prism<(A, R)> for Z { type Elem = A; ... };
impl<N, A, R> Prism<(A, R)> for S<N> where N: Prism<R> { type Elem = N::Elem; ... }
\end{lstlisting}

The \code{Prism} trait is parameterized over a row type
\code{Row=HList![...]}, with the associated type \code{Elem} being the
element type that has been selected from the list by the prism. We
then inductively implement \code{Prism} using type-level natural
numbers, with the number \code{N} used for selecting the N-th element
of the heterogeneous list.  The definition of \code{Prism} is similar
to \code{ContextLens}, with the main difference being that we only
need \code{Prism} to support extraction and injections operations on
the sum types that are derived from the heterogeneous list.
Using \code{Prism}, a generalized \code{offer\_case} function is implemented as follows:

\begin{lstlisting}[language=Rust, style=nicerust]
fn offer_case<C: Context, A: Protocol, Row, N: Prism<Row, Elem=A>>
  (n: N, cont: PartialSession<C, A>) -> PartialSession<C, InternalChoice<Row>>
\end{lstlisting}

The function accepts a natural number \code{N} as the first parameter,
which acts as the \textit{prism} for selecting a session type $A_N$ out of the
row type \codee{Row=HList![..., A$_{\texttt{N}}$, ...]}. Through the associated type \code{A=N::Elem},
\code{offer\_case} forces the programmer to provide a continuation that offers the chosen
session type \code{A}.

\subsection{Binary Branching}

While \code{offer\_case} is a step in the right direction, it only
allows the selection of a specific choice, but not the provision of
\emph{all} possible choices.  The latter, however, is necessary to
encode the $\sillr$ left rule of internal choice and right rule of
external choice.  To illustrate the problem, let's consider the right
rule of a binary external choice, $\textsc{T$\extchoice\m{2}_\m{R}$}$:

\vspace{-10pt}
\begin{center}
  \begin{minipage}[c]{0.5\textwidth}
    {\small
\[
\inferrule*[right=($\textsc{T$\extchoice\m{2}_\m{R}$}$)]
{ \Gamma \, ; \, \Delta \entails K_l :: A
  \\
  \Gamma \, ; \, \Delta \entails K_r :: B
}
{ \Gamma \, ; \, \Delta \entails
  \m{offer\_choice\_2} \; K_l \; K_r :: A \extchoice B
}
\]}
\end{minipage}
\end{center}
\vspace{-10pt}

The \code{offer\_choice\_2} construct has two possible continuations
$K_l$ and $K_r$, with only one of them being executed, depending on
the selection by the client.  In a naive implementation, we can define
the construct to accept two continuations as follows:

\begin{lstlisting}[language=Rust, style=nicerust]
fn offer_choice_2<C: Context, A: Protocol, B: Protocol>
  ( cont_left: PartialSession<C, A>, cont_right: PartialSession<C, B> )
  -> PartialSession<C, ExternalChoice2<A, B>>
\end{lstlisting}

While the above implementation works in most languages, it is not
adequate in Rust.  Since Rust's type system is \textit{affine},
variables can only be captured by one of the continuation closures,
but not both. As far as the compiler is aware, both closures can
potentially be called, and we cannot state that one of
the branches is guaranteed to never run.

In order for \code{offer\_choice\_2} to work in Rust's affine typing,
it has to accept only one continuation closure and have it return
either \code{PartialSession<C, A>} or \code{PartialSession<C, B>},
depending on the client's selection. It is not as straightforward to
express such behavior as a valid type in a language like Rust. If Rust
supported dependent types, \code{offer\_choice\_2} could be
implemented along the following lines:

\begin{lstlisting}[language=Rust, style=nicerust]
fn offer_choice_2<C: Context, A: Protocol, B: Protocol>
  ( cont: impl FnOnce(first: bool) ->
      if first { PartialSession<C, A> } else { PartialSession<C, B> } )
  -> PartialSession<C, ExternalChoice2<A, B>>
\end{lstlisting}

\noindent That is, the return type of the \code{cont} closure depends
on the whether the \textit{value} of the \code{first} argument is true
or false. However, since Rust does not support dependent types, we
emulate a dependent sum in a non-dependent language, using a CPS
transformation:

\begin{lstlisting}[language=Rust, style=nicerust]
fn offer_choice_2<C: Context, A: Protocol, B: Protocol>
  ( cont: impl FnOnce(InjectSum2<C, A, B>) -> ContSum2<C, A, B> )
    -> PartialSession<C, ExternalChoice2<A, B>>
\end{lstlisting}

The function \code{offer\_choice\_2} accepts a continuation function
\code{cont} that is given a value of type \code{InjectSum2<C, A, B>}
and returns a value of type \code{ContSum2<C, A, B>}. We will now look
at the definitions of \code{ContSum2} and \code{InjectSum2}.  First,
we observe that the different return types for the two branches can be
unified with a type \code{ContSum2}:

\begin{lstlisting}[language=Rust, style=nicerust]
struct ContSum2<C: Context, A: Protocol, B: Protocol> { ... }
async fn run_cont_sum<C: Context, A: Protocol, B: Protocol>(cont: ContSum2<C, A, B>)
\end{lstlisting}

The type \code{ContSum2} contains the necessary data for executing
either a \code{PartialSession<C, A>} or a \code{PartialSession<C, B>},
together with the runtime data for the linear context \code{C}. For
brevity, the implementation details of \code{ContSum2} are omitted,
with the private function \code{run\_cont\_sum} provided as an abstraction for
Ferrite to execute the continuation.

We then define \code{InjectSum2} as a sum of boxed closures that would
construct a \code{ContSum2} from either a \code{PartialSession<C, A>}
or a \code{PartialSession<C, B>}:

\begin{lstlisting}[language=Rust, style=nicerust]
enum InjectSum2<C, A, B> {
  InjectLeft(Box<dyn FnOnce(PartialSession<C, A>) -> ContSum2<C, A, B>>),
  InjectRight(Box<dyn FnOnce(PartialSession<C, B>) -> ContSum2<C, A, B>>) }
\end{lstlisting}

When the \code{cont} passed to \code{offer_choice_2} is given a value
of type \code{InjectSum2<C, A, B>}, it has to branch on it and match
on whether the \code{InjectLeft} or \code{InjectRight} constructors
are used. Since the return type of \code{cont} is \code{ContSum2<C, A,
  B>} and the constructor for \code{ContSum2} is private, there is no
other way for \code{cont} to construct the return value other than to
call either \code{InjectLeft} or \code{InjectRight} with the
appropriate continuation.

The use of \code{InjectSum2} prevents the programmer from providing
the wrong branch in the continuation by keeping the constructor
private.  However a private constructor alone cannot prevent two uses
of \code{InjectSum2} to be \textit{deliberately} interchanged, causing
a protocol violation.  To fully ensure that there is no way for the
user to provide a \code{ContSum2} from elsewhere, we instead use a
technique from GhostCell~\cite{GhostCell21} that uses
\textit{higher-ranked trait bounds} (HTRB) to mark a phantom invariant
lifetime on both \code{InjectSum2} and \code{ContSum2}:

\begin{lstlisting}[language=Rust, style=nicerust]
fn offer_choice_2<C: Context, A: Protocol, B: Protocol>
  ( cont: for <'r> impl FnOnce(InjectSum2<'r, C, A, B>) -> ContSum2<'r, C, A, B> )
    -> PartialSession<C, ExternalChoice2<A, B>>
\end{lstlisting}

The use of HRTB ensures that each call of \code{offer\_choice\_2}
would generate a unique lifetime \code{'r} for the continuation.
Using that, Ferrite can ensure that a value of type
\code{InjectSum2<'r1, C, A, B>} cannot be used to construct the
return value of type \code{ContSum2<'r2, C, A, B>}, if the lifetimes
\code{<'r1>} and \code{<'r2>} are different.  An example use of
\code{offer\_choice\_2} is as follows:

\begin{lstlisting}[language=Rust, style=nicerust]
let choice_provider: Session<ExternalChoice2<SendValue<u64, End>, End>>
  = offer_choice_2(| b | { match b {
        InjectLeft(ret) => ret(send_value(42, terminate())),
        InjectRight(ret) => ret(terminate()) } });
\end{lstlisting}

The example code above requires some boilerplate code to call the
session injector \code{ret} to wrap around the continuation
expression.  To free the programmer from writing such boilerplate,
Ferrite also provides a macro \code{offer\_choice} that translates
into the underlying pattern matching syntax, which is explained in the
next section.

\subsection{N-ary Branching}

To generalize \code{offer\_choice\_2} to n-ary choices, Ferrite has
its own version of polymorphic variants implemented in Rust. Our
implementation specifically targets Rust, and is based on similar work
by \cite{Morris15} and \cite{RowsMorrisM19}.  The base variant types
are as follows:

\begin{lstlisting}[language=Rust, style=nicerust]
enum Bottom {}      enum Sum<A, B> { Inl(A), Inr(B) }
trait TypeApp<A> { type Applied; }      trait SumApp<F> { type Applied; }
type App<F, A> = <F as TypeApp<A>>::Applied;
type AppSum<Row, F> = <Row as SumApp<F>>::Applied;
impl<F> SumApp<F> for () { type Applied = Bottom; }
impl<A, F: TypeApp<A>, R: SumApp<F>> SumApp<F> for (A, R) {
  type Applied = Sum<F::Applied, R::Applied>; }
\end{lstlisting}

Similar to \code{RecApp} described in \Cref{sec:recursive-types},
\code{TypeApp} is used to represent a Rust type emulating the kind
$\m{Type} \rightarrow \m{Type}$ for non-recursive usage.  Furthermore,
the \code{SumApp} trait is used to represent a Rust type emulating the
kind $(\m{Type} \rightarrow \m{Type}) \rightarrow \m{Type}$.  The type
alias \code{App<F, A>} is used to extract the associated type
\code{Applied} when \code{F} is applied to \code{A} via
\code{TypeApp}.  The type alias \code{AppSum<Row, F>} is used to
extract the associated type \code{Applied} when a row type \code{Row}
is applied to a type constructor \code{F}, which implements
\code{TypeApp<A>} for all \code{A}.
\footnote{ For brievity, we omit some
  details that the types \texttt{App} and \texttt{AppSum} are actually
  implemented as structs that are \textit{isomorphic} to \texttt{<F as
    TypeApp<A>{}>::Applied} and \texttt{<Row as
    SumApp<F>{}>::Applied}, resp.  The main difference is that
  the actual structs \textit{hide} the propagation of the trait bound
  requirements of \texttt{TypeApp} and \texttt{SumApp} from their
  callers, resulting in much cleaner code.  This does not affect the
  understanding of the core concepts introduced in this section.  }

Using \code{SumApp}, we map an heterogeneous list to nested sums such
that \code{AppSum<HList![A0, A1, ...], F> = Sum![App<F, A0>, App<F,
  A1>, ...]}, with the macro \code{Sum!} used to expand the macro
arguments into nested sums, \ie \code{Sum![A0, A1, ...] = Sum<A0,
  Sum<A1, ..., Bottom>>}.
We  then define the n-ary versions of \code{InjectSum2} and
\code{ContSum2} as follows:

\begin{lstlisting}[language=Rust, style=nicerust]
struct InjectSessionF<'r, Row, C> {}      struct InjectSession<'r, Row, C, A> { ... }
struct ContSum<'r, Row, C: Context> { ... }
impl<'r, Row, C: Context> TypeApp<A> for InjectSessionF<C> {
  type Applied = InjectSession<C, A>; }
impl<'r, Row, C: Context, A: Protocol> FnOnce(PartialSession<C, A>)
  -> ContSum<'r, Row, C> for InjectSession<'r, Row, C, A> { ... }
\end{lstlisting}

The type \code{InjectSessionF<'r, Row, C>} serves as a marker type for
\code{TypeApp}, such that when applied to a type \code{A}, we get the
struct \code{InjectSession<'r, Row, C, A>}. Conceptually, the struct
implements the trait \code{FnOnce(PartialSession<C, A>) -> ContSum<'r,
  Row, C>}, so that we can apply a \code{PartialSession<C, A>} to it
and get back a \code{ContSum<'r, Row, C>}.\footnote{Technically, Rust
  does not allow custom implementation of \code{FnOnce}, so Ferrite
  defines a custom trait with the same behavior.}
The composed type \code{AppSum<Row, InjectSessionF<'r, Row, C>}
represents a row of \code{InjectSession}, with \code{Row} being a
heterogeneous list in the form \codee{HList![A$_0$, A$_1$, ...,
  A$_{N-1}$]}.  For example, the type \code{AppSum<HList![A, B],
  InjectSessionF<'r, Row, C>>} evaluates to
\code{Sum![InjectSession<'r, Row, C, A>, InjectSession<'r, Row, C,
  B>]}, which is isomorphic to the type \code{InjectSum2<C, A, B>}
that we defined for the binary case.  Using the row constructs, we can
define n-ary version \code{offer\_choice} as follows:

\begin{lstlisting}[language=Rust, style=nicerust]
fn offer_choice<C: Context, Row>(cont1 : impl for <'r>
  FnOnce(AppSum<Row, InjectSessionF<'r, Row, C>>) -> ContSum<'r, Row, C>
) -> PartialSession<C, ExternalChoice<Row>>
where Row: SumApp<InjectSessionF<'r, Row, C>>>, ...
\end{lstlisting}

With the n-ary version of \code{offer\_choice} available, we can
re-implement binary choice as a specialized version.  To do that, we
only need a few type aliases and struct definitions to make the syntax
more pleasing:

\begin{lstlisting}[language=Rust, style=nicerust]
enum EitherSum<A, B> { Left(A), Right(B) };     type Either<A, B> = HList![A, B];
const LeftLabel: Z = Z::new();                  const RightLabel: S<Z> = <S<Z>>::new();
impl<A, B> std::convert::From<Sum![A, B]> for EitherSum<A, B> { ... }
\end{lstlisting}

We first define an \code{EitherSum} enum, and a
\code{std::convert::From} instance that converts an unlabeled nested
sum \code{Sum![A, B]} into the labeled sum \code{EitherSum<A, B>}.
The conversion allows users to use a flat list of labeled match arms
during branching, and give meaningful labels \code{Left} and
\code{Right} to each branch.
We also define \code{Either<A, B>} as a type alias to the row type
\code{HList![A, B]}, to give a meaningful name to the choice
protocol. Finally we define the constants \code{LeftLabel} and
\code{RightLabel} to refer to the prisms \code{Z} and \code{S<Z>},
resp.  Ferrite also provides a helper macro
\code{define\_choice!} to help users define custom choice protocols
that look similar to the above. This is used in conjunction with
macros such as \code{offer\_choice!}, which cleans up the boilerplate
required to enable different match branches to return different types.
Using the macros, users can define the same \code{Either} protocol and
write an external choice provider as follows:

\begin{lstlisting}[language=Rust, style=nicerust]
define_choice!{ Either<A, B>; Left: A, Right: B }
// Inferred type: Session<ExternalChoice<Either<SendValue<u64, End>, End>>>
let provider = offer_choice!{
  Left => send_value(42, terminate()), Right => terminate() };
\end{lstlisting}

For convenience, Ferrite exports the choice definition for \code{Either}
for the anonymous declaration of binary choice in session types.


\section{Evaluation}\label{sec:evaluation}

\definecolor{servo-color}{RGB}{0, 36, 180}
\definecolor{ferrite-color}{RGB}{36, 80, 0}

\lstdefinestyle{servo}{
  commentstyle=\color{servo-color},
  numberstyle=\color{servo-color},
  basicstyle=\footnotesize\ttfamily\color{servo-color},
  keywordstyle= \bfseries\color{servo-color},
}

\lstdefinestyle{servo-ferrite}{
  basicstyle=\footnotesize\ttfamily\color{ferrite-color},
  keywordstyle=\bfseries\color{ferrite-color},
}

The Ferrite library is more than just a research prototype. It is
designed for practical use in real world applications. To evaluate the
design and implementation of Ferrite, we re-implemented the
communication layer of the canvas component of
Servo~\cite{ServoWebsite} entirely in Ferrite.  Servo is an under
development browser engine that uses message-passing for heavy task
parallelization. Canvas provides 2D graphic rendering,
allowing clients to create new canvases and perform operations on
a canvas such as moving the cursor and drawing shapes.

The canvas component is a good target for evaluation as it is
sufficiently complex and also very demanding in terms of performance.
Canvas is commonly used for animations in web applications.  For an
animation to look smooth, a canvas must render at least 24 frames per
second, with potentially thousands of operations to be executed per
frame.

The changes we made are fairly minimal, consisting of roughly 750 lines of
additions and 620 lines of deletions, out of roughly
300,000 lines of Rust code in Servo.  The sources of our
implementation are provided as an artifact.  To
differentiate the two versions of code snippets, we use
{\color{servo-color}blue} for the original code, and
{\color{ferrite-color}green} for the code using Ferrite.

\subsection{Servo Canvas Component}

\begin{figure}
\centering
\input{code/servo}
\vspace{-10pt}
\caption{Message-passing concurrency in Servo's canvas component (simplified for
  illustration purposes).}
\vspace{-15pt}
\label{code:servo}
\end{figure}

\Cref{code:servo} provides a sketch of the main communication paths in
Servo's canvas component~\cite{ServoCanvas}.  The canvas component is
implemented by the \code{CanvasPaintThread}, whose function
\code{start} contains the main communication loop running in a
separate thread
(lines~\ref{code:fn-start-loop-start}--\ref{code:fn-start-loop-end}).
This loop processes client requests received along
\code{canvas\_msg\_receiver} and \code{create\_receiver}, which are
the receiving endpoints of the channels created prior to spawning the
loop
(lines~\ref{code:fn-start-chn-canvas}--\ref{code:fn-start-chn-constellation}).
The channels are typed with the enumerations
\code{ConstellationCanvasMsg} and \code{CanvasMsg}, defining messages
for creating and terminating the canvas component and for executing
operations on an individual canvas, resp.
When a client sends a message that expects a response from the
recipient, such as \code{GetTransform} and \code{IsPointInPath}
(lines~\ref{code:get-transform}--\ref{code:is-point-in-path}), it
sends a channel along with the message to be used by the recipient to
send back the result.
Canvases are identified by an id, which is
generated upon canvas creation (line~\ref{code:fn-start-id-creation})
and stored in the thread's \code{canvases} hash map
(line~\ref{code:canvases-hashmap}).  If a client requests an
invalid id, for example after prior termination and removal of the
canvas (line~\ref{code:fn-start-canvas-close}), the failed assertion
\code{expect("Bogus canvas id")} (line~\ref{code:fn-start-bogusid})
will result in a \code{panic!}, causing the canvas component to crash
and subsequent calls to fail.

The code in \Cref{code:servo} uses a clever combination of
enumerations to type channels and ownership to rule out races on the
data sent along channels.  Nonetheless, Rust's type system is not
expressive enough to enforce the intended \emph{protocol} of message
exchange and existence of a communication partner.  The latter is a
consequence of Rust's type system being \emph{affine}, which permits
``dropping of a resource''.  The dropping or premature closure of a
channel, however, can result in a proliferation of \code{panic!} and
thus cause an entire application to crash.  In fact, while refactoring
Servo to use Ferrite, we were able to uncover a protocol violation in
Servo, caused by one of the nested match arms of the provider doing an
early return before sending back any result to the client.

\subsection{Canvas Protocol in Ferrite}

In the original canvas component, the provider \code{CanvasPaintThread} accepts
messages of type \code{CanvasMsg}, made up of a combination of smaller
sub-message types such as \code{Canvas2dMsg}. We note that the majority
of the sub-message types have the following trivial form:

\begin{lstlisting}[language=Rust, style=servo]
enum CanvasMsg { Canvas2d(Canvas2dMsg, CanvasId), Close(CanvasId), ... }
enum Canvas2dMsg { BeginPath, ClosePath, Fill(FillOrStrokeStyle), ... }
\end{lstlisting}

The trivial sub-message types such as \code{BeginPath}, \code{Fill},
and \code{LineTo} do not require a response from the provider, so the
client can simply fire them and proceed. Although we can offer all
sub-message types as separate branches in an external choice, it is
more efficient to keep trivial sub-messages in a single enum. In
our implementation, we define \code{CanvasMessage} to have
similar sub-messages as \code{Canvas2dMsg}, with non-trivial messages
such as \code{IsPointInPath} moved to separate branches.

\begin{lstlisting}[language=Rust, style=servo-ferrite]
enum CanvasMessage { BeginPath, ClosePath, Fill(FillOrStrokeStyle), ... }
define_choice! { CanvasOps; Message: ReceiveValue<CanvasMessage, Release>, ... }
type Canvas = LinearToShared<ExternalChoice<CanvasOps>>;
\end{lstlisting}

We use the \code{define\_choice!} macro described in
\Cref{sec:choices} to define an n-ary choice \code{CanvasOps}. The
first branch of \code{CanvasOps} is labelled \code{Message}, and the
only action is for the provider to receive a \code{CanvasMessage}. The
choices are offered as an external choice, and the session type
\code{CanvasProtocol} is defined as a shared protocol that offers the
choices in the critical section.

The original design of the \code{CanvasPaintThread} would be
sufficient if the only messages being sent were trivial messages.
However, \code{Canvas2dMsg} also contains non-trivial sub-messages,
such as \code{GetImageData} and \code{IsPointInPath}, demanding a
response from the provider:

\begin{lstlisting}[language=Rust, style=servo]
enum Canvas2dMsg { ..., GetImageData(Rect<u64>, Size2D<u64>, IpcBytesSender),
  IsPointInPath(f64, f64, FillRule, IpcSender<bool>), ... }
\end{lstlisting}

To obtain the result from the original canvas, clients must create a
new inter-process communication (IPC) channel and bundle the channel's
sender
endpoint with the message. In our
implementation, we define separate branches in
\code{CanvasOps} to handle non-trivial cases:

\begin{lstlisting}[language=Rust, style=servo-ferrite]
define_choice! { CanvasOps; Message: ReceiveValue<CanvasMessage, Release>,
  GetImageData: ReceiveValue<(Rect<u64>, Size2D<u64>), SendValue<ByteBuf, Release>>,
  IsPointInPath: ReceiveValue<(f64, f64, FillRule), SendValue<bool, Release>>, ... }
\end{lstlisting}

The original \code{GetImageData} accepts an \code{IpcBytesSender},
which sends raw bytes back to the client. In Ferrite, we translate the
use of \code{IpcBytesSender} to the type
\code{SendValue<ByteBuf, Z>}, which sends the raw bytes wrapped in a
\code{ByteBuf} type. We discuss possible performance penalties of
this approach in \Cref{sec:benchmark}.

Aside from the \code{Canvas} protocol, we also redesign the use of
\code{ConstellationCanvasMsg} into its own shared protocol,
\code{ConstellationCanvas}:

\begin{lstlisting}[language=Rust, style=servo-ferrite]
type ConstellationCanvas = LinearToShared<ReceiveValue<Size2D,
    SendValue<SharedChannel<Canvas>, Release>>>;
\end{lstlisting}

To create a new canvas, a client first acquires the shared channel of
type \code{SharedChannel<ConstellationCanvas>}. Afterwards, the client
sends the \code{Size2D} parameter to specify the canvas size. The
constellation canvas provider then spawns a new canvas shared process
through \code{run\_shared\_session} and sends back the shared channel
of type \code{SharedChannel<Canvas>} as a value. Finally, the session
is released, allowing other clients to acquire the shared provider.

\subsection{Performance Evaluation}\label{sec:benchmark}

To evaluate the performance of the canvas component, we use the
MotionMark benchmark suite \cite{MotionMark}. MotionMark is a web
benchmark that focuses on graphics performance of web browsers.  It
contains benchmarks for various web components, including canvas, CSS,
and SVG.  As MotionMark does not yet support Servo, we modified the
benchmark code to make it work in the absence of features that are not
implemented in Servo.

We provide the modified benchmark source code
along with instructions for running it as an artifact.
\Cref{sec:servo-challenges} is also provided to highlight
some of the implementation challenges in porting Servo to use
Ferrite, in particular on the latency incurred by inter-process
communication, and our workaround to compensate the complication.

For the purpose of this evaluation, we focused on benchmarks that
target the canvas component and skipped benchmarks that fail in Servo
due to missing features. We ran each benchmark in a fixed 1600x800
resolution for 30 seconds, on a Core i7 Linux desktop machine. We ran
the benchmarks against the original Servo, modified Servo with Ferrite
canvas (Servo/Ferrite), Firefox, and Chrome.  Our performance scores
are measured in the fixed mode version of MotionMark, which measures
frames per second (fps) performance of executing the same set of
canvas operations per frame.

\begin{table*}[t]
\caption{MotionMark Benchmark scores in fps (higher is better)}
\vspace{-10pt}
\label{tab:benchmark-results}
\begin{footnotesize}
\begin{center}
\begin{tabular}{|l |c| c| c| c |}
\hline
\textbf{Benchmark Name} 
  & \textbf{Servo} &
\textbf{Servo/Ferrite} & \textbf{Firefox} & \textbf{Chrome} \\ [0.5ex]
\hline
Arcs                        
  & 12.21 $\pm$ 6.75\%  & 11.83 $\pm$ 11.49\%  & 52.61 $\pm$ 32.88\%  & 46.00 $\pm$ 9.00\%  \\
Paths                       
  & 43.76 $\pm$ 10.66\%  & 40.98 $\pm$ 18.94\%  & 55.59 $\pm$ 28.80\%  & 59.50 $\pm$ 14.90\%  \\
Lines                       
  &  7.48 $\pm$ 7.06\%  & 11.47 $\pm$ 12.74\%  & 14.35 $\pm$ 6.65\%  & 32.43 $\pm$ 6.48\%  \\
Bouncing clipped rects      
  & 18.43 $\pm$ 7.06\%  & 18.23 $\pm$ 11.00\%    & 34.82 $\pm$ 7.76\%  & 58.07 $\pm$ 19.85\% \\
Bouncing gradient circles   
  &  8.02 $\pm$ 7.74\%  &  7.72 $\pm$ 12.63\%    & 58.79 $\pm$ 21.03\%  & 59.77 $\pm$ 10.07\% \\
Bouncing PNG images         
  &  7.97 $\pm$ 5.91\%  &  6.31 $\pm$ 10.26\%    & 24.61 $\pm$ 6.35\%  & 59.94 $\pm$ 13.04\% \\
Stroke shapes               
  &  10.60 $\pm$ 3.95\%  &  10.35 $\pm$ 10.96\%    & 51.21 $\pm$ 11.25\%  & 59.38 $\pm$ 16.87\% \\
Put/get image data          
  & 60.01 $\pm$ 3.81\%  & 32.08 $\pm$ 10.83\%    & 59.66 $\pm$ 20.16\%  & 60.00 $\pm$ 5.00\% \\
\hline
\end{tabular}
\end{center}
\end{footnotesize}
\vspace{-10pt}
\end{table*}

The benchmark results are shown in \Cref{tab:benchmark-results},
with the performance scores in fps (higher fps is better).
It is worth noting that a benchmark can achieve at most 60 fps.
Our goal in this benchmark is to keep the scores of Servo/Ferrite close
to those of Servo, \textit{not} to achieve better performance than the original.
This is shown to be the case in most of the benchmarks.

The only benchmark with a large difference between Servo and
Servo/Ferrite is \textit{Put/get image data}, with Ferrite performing
2x worse.  This is because in Servo/Ferrite, we use \code{ByteBuf}
to transfer the images as raw bytes within the same shared channel.
In contrast, Servo uses a specialized structure \code{IpcBytesSender}
for transferring of raw bytes in parallel to other messages. As a result,
the communication in Servo/Ferrite is congested during the transfer of
the image data, while the original Servo can process new messages
in parallel to the image data being transmitted.

We also observe that there are significant performance differences in
the scores between Servo and those in Firefox and Chrome, indicating
that there exist performance bottlenecks in Servo unrelated to
communication protocols.


\section{Related and Future Work}\label{sec:related}






Session type embeddings exist for various languages, including
Haskell~\cite{PucellaHaskell2008, ImaiPLACES2010, LindleyHaskell2016,
  OrchardPOPL2016}, OCaml~\cite{Padovani17, ImaiARTICLE2019},
Java~\cite{HuECOOP2008,HuECOOP2010}, and Scala~\cite{ScalasECOOP2016}.
Functional languages like ML, OCaml, and Haskell, in particular, are
ideal host languages for creating EDSLs thanks to their advanced
features (e.g.~type classes, type families, higher-rank and
higher-kinded types and GADTs).
\cite{PucellaHaskell2008} first demonstrated the feasibility of
embedding session types in Haskell, with refinements done in later works
\cite{ImaiPLACES2010, LindleyHaskell2016, OrchardPOPL2016}.
Similar embeddings have also been contributed
in the context of OCaml by \code{FuSe} \cite{Padovani17} and
\code{session-ocaml} \cite{ImaiARTICLE2019}.

Aside from Ferrite, there are other implementations of session types
in Rust, including \code{session\_types}~\cite{JespersenWGP2015},
\code{sesh}~\cite{KokkeICE2019}, and
\code{rumpsteak}~\cite{Rumpsteak2021Coord, Rumpsteak2021Arxiv}.
\code{session\_types} were the first implementation to make use of
affinity to provide a session type library in Rust.  \code{sesh}
emphasizes this aspect by embedding the affine session type system
Exceptional GV~\cite{FowlerPOPL2019} in Rust. Both
\code{session\_types} and \code{sesh} adopt a classical perspective,
requiring the endpoints of a channel to be typed with dual types.
\code{rumpsteak} develops an embedding of multiparty session types by
generating Rust types derived from multiparty session types defined in
Scribble~\cite{Scribble2013}.

Due to their reliance on Rust's affine type system, neither
\code{session\_types} nor \code{sesh} prevents a channel endpoint from
being dropped prematurely, relegating the handling of such errors to
the runtime. \code{rumpsteak} uses some type-level techniques similar
to Ferrite to enforce a channel's linear usage in the continuation
passed to the \code{try\_session} function.  This ensures that a
linear channel in \code{rumpsteak} is always fully consumed, if it is
ever consumed. However, prior to the call to \code{try\_session},
the linear channel exist as an affine value, which may be dropped by
the Rust program without being consumed at all, thereby causing
deadlock. In comparison, Ferrite enforces linearity at all level,
including safe linking of multiple linear processes using \code{cut}.

In terms of concurrency, \code{session\_types}, \code{sesh}, and
\code{rumpsteak} all require the programmer to manually manage
concurrency, either by spawning threads or async tasks.  This
introduces potential failure when the code fails follow the requirement
to spawn all processes. On the other hand, the simplicity of such a model
allows relatively few threads or async tasks to be spawned, thereby allowing
the underlying runtime to execute the processes more efficiently.  In
comparison, Ferrite offers fully managed concurrency, without the
programmer having to worry about how to spawn the processes and
execute them in parallel.

In terms of performance, the downside of Ferrite's concurrency
approach is that it aggresively spawns new async tasks in each use of
\code{cut}.  Although async tasks in Rust are much more lightweight
than OS threads, there is still a significant overhead in spawning and
managing many async tasks, especially in micro-benchmarks.  As a
result, Ferrite tends to perform slower than alternative Rust
implementations in settings where only a fixed small number of
processes need to be spawned. Nevertheless, it is worth noting that
the async ecosystem in Rust is still relatively immature, with many
potential improvements to be made. In practice, the overhead of the
async runtime may also be negligible when compared to the core
application logic.  In such cases, Ferrite would also allow
applications to scale more easily by allowing many more processes to
be spawned and managed concurrently without requiring additional
effort from the programmer.

In terms of DSL design, Ferrite is more closely related to the
embeddings in OCaml and Haskell, as it fully enforces a linear
treatment of session type channels and thus \emph{statically} rules
out any panics arising from dropping a channel prematurely. Ferrite
also differs from other libraries in that it adopts intuitionistic
typing~\cite{CairesCONCUR2010}, allowing the typing of a channel
rather than its two endpoints.  On the use of profunctor optics, our
work is the first to connect n-ary choice to prisms, while prior work
by \code{session-ocaml} \cite{ImaiPLACES2010} has only established the connection between
lenses, the dual of prisms, and linear contexts.  \code{FuSe} \cite{Padovani17}
and \code{session-ocaml} \cite{ImaiARTICLE2019} have previously explored the use of n-ary
(generalized) choice through extensible variants available only in
OCaml. Our work demonstrates that it is possible to encode extensible
variants, and thus n-ary choice, as type-level constructs using
features available in Rust.

A major difference in terms of implementation is that Ferrite uses a
continuation-passing style, whereas Haskell and OCaml embeddings
commonly use (indexed) monads and do-notation style.  This technical
difference amounts to a key conceptual one: a direct correspondence
between the Rust programs generated from Ferrite constructs and the
\SILLR typing derivation.  As a result, the generated Rust code can be
viewed as carrying the proof of protocol adherence.

The embeddings of \code{ESJ} \cite{HuECOOP2010} and \code{lchannels} \cite{ScalasECOOP2016} also
adopt a continuation-passing style, but do not faithfully embed typing
derivations (i.e.~they do not statically enforce linearity).  These
approaches follow an encoding of session types using linear
types~\cite{DBLP:conf/ppdp/DardhaGS12} first proposed by
Kobayashi~\cite{DBLP:conf/unu/Kobayashi02} in the setting of
$\pi$-calculus. Type systems for message-passing in $\pi$-calculus
have a long history, dating back to the work of Kobayashi and
Igarashi~\cite{DBLP:conf/sas/IgarashiK97,DBLP:conf/popl/IgarashiK01,DBLP:journals/tcs/IgarashiK04}.
These systems often focus on (but are not limited to) deadlock-freedom
and lock-freedom~\cite{DBLP:journals/toplas/KobayashiS10} by enforcing
a partial order on matching communication. This approach has been
studied for the linear $\pi$-calculus~\cite{DBLP:conf/csl/Padovani14}
and in the presence of interrupts~\cite{DBLP:conf/esop/SuenagaK07} or
unbounded process networks~\cite{DBLP:conf/concur/GiachinoKL14}.
While session types are generally less powerful than the approaches of
Kobayashi et al., they provide a useful compromise between
expressiveness and simplicity, being more amenable to embeddings in
general-purpose language constructs and type systems.

In terms of expressiveness, Ferrite contributes over all prior
session-based works in its support for shared session
types~\cite{BalzerICFP2017}, allowing it to express real-world
protocols, as demonstrated by our implementation of Servo's canvas
component. Shared session types reclaim the expressiveness of the
untyped asynchronous $\pi$-calculus in session-typed
languages~\cite{BalzerCONCUR2018}, at the cost of deadlock-freedom.
Recent extensions of classical linear logic session types contribute
another approach to softening the rigidity of linear session types to
support multiple client sessions and
nondeterminism~\cite{QianARXIV2020} and memory cells and
nondeterministic updates~\cite{DBLP:journals/pacmpl/RochaC21}, resp.

Our technique of a judgmental embedding opens up new possibilities for
embedding type systems other than session types in Rust.  Although we
have demonstrated that the judgmental embedding is sufficiently
powerful to encode a type system like session types, the embedding is
currently \textit{shallow}, with the implementation hardcoded to use
the channels and async run-time from \code{tokio}.  Rust comes with unique
features such as affine types and lifetimes that makes it especially
suited for implementing concurrency primitives, as evidenced by the
wealth of channel and async run-time implementations
available.  As discussed in \Cref{sec:evaluation}, one of our future
goals is to explore the possibility of making Ferrite a \textit{deep}
embedding of session types in Rust, so that users can choose from
multiple low-level implementations.  Although deep embeddings have
extensively been explored for languages like Haskell
\cite{SvenningssonA12, LindleyHaskell2016}, it remains a open question
to find suitable approaches that work well in Rust.


\bibliography{references}

\newpage

\appendix

\section{Typing Rules}\label{sec:typing-rules}

\begin{table*}
\caption{Overview of session terms in $\sillr$ and Ferrite.}
\label{tab:term-mapping}
\centering
\begin{scriptsize}
\renewcommand{\tabcolsep}{2mm}
\begin{tabular}{@{}lp{3cm}p{4.8cm}p{4.5cm}@{}}
\toprule
\textbf{\SILLR} &
\textbf{Ferrite} &
\textbf{Term} &
\textbf{Description} \\
\midrule
$\mb{\epsilon}$ &
\texttt{End} &
$\m{terminate};$ &
Terminate session. \\
&
&
$\m{wait} \; a; \, K$ &
Wait for channel $a$ to close. \\\\
$\mi{\tau} \recvval A$ &
\texttt{ReceiveValue<T, A>} &
$x \leftarrow \m{receive\_value}; \, K$ &
Receive value $x$ of type $\tau$. \\

&
&
$\m{send\_value\_to} \; a \; x; \, K$ &
Send value $x$ of type $\tau$ to $a$. \\
$\mi{\tau} \sendval A$ &
\texttt{SendValue<T, A>} &
$\m{send\_value} \; x; \, K$ &
Send value of type $\tau$. \\

&
&
$x \leftarrow \m{receive\_value\_from} \, a \, x; \, K$ &
Receive value of type $\tau$ from channel $a$. \\
$A \chanin B$ &
\texttt{ReceiveChannel<A, B>} &
$a \leftarrow \m{receive\_channel}; \, K$ &
Receive channel $a$ of session type $A$. \\

&
&
$\m{send\_channel\_to} \; f \; a; \, K$ &
Send channel $a$ to channel $f$ of session type $A \chanin B$. \\
$A \chanout B$ &
\texttt{SendChannel<A, B>} &
$\m{send\_channel\_from} \; a; \, K$ &
Send channel $a$ of session type $A$. \\

&
&
$a \leftarrow \m{receive\_channel\_from} \; f \; a; \, K$ &
Receive channel $a$ from channel $f$ of session type $A \chanout B$. \\

$\lineartoshared A$ &
\texttt{LinearToShared<A>} &
$\m{accept\_shared\_session}; \, K_l$ &
Accept an acquire, then continue as linear session $K_l$. \\

&
&
$a \leftarrow \m{acquire\_shared\_session} \, s; \, K_l$ &
Acquire shared channel $s$ as linear channel $a$. \\
$\sharedtolinear S$ &
\texttt{SharedToLinear<S>} &
$\m{detach\_shared\_session}; \, K_s$ &
Detach linear session and continue as shared session $K_s$. \\

&
&
$\m{release\_shared\_session} \, a; \, K_l$ &
Release acquired linear session. \\

$A \extchoice B$ &
\texttt{ExternalChoice< Either<A, B>{}>} &
$\m{offer\_choice\_2} \; K_l \; K_r$ &
Offer either continuation $K_l$ or $K_r$ based on client's choice. \\

&
&
$\m{choose\_left} \; a; \, K$ &
Choose the left branch offered by channel $a$  \\

&
&
$\m{choose\_right} \, a; \, K$ &
Choose the right branch offered by channel $a$  \\

$A \intchoice B$ &
\texttt{InternalChoice< Either<A, B>{}>} &
$\m{offer\_left}; \, K$ &
Offer the left branch \\

&
&
$\m{offer\_right}; \, K$ &
Offer the right branch \\

&
&
$\m{case\_2} \; a \; K_l \; K_r$ &
Branch to either $K_l$ or $K_r$ based on choice offered by channel $a$. \\

$\extchoice\{ \overline{ l_i: A_i } \}$ &
\texttt{ExternalChoice<Row>} &
$\m{offer\_choice} \{ \overline{ l_i: K_i } \} $ &
Offer continuation $K_i$ when the client selects $l_i$.\\

&
&
$\m{choose} \; a \; l_i ; \, K$ &
Choose the $l_i$ branch offered by channel $a$  \\

$\intchoice\{ \overline{ l_i: A_i } \}$ &
\texttt{InternalChoice<Row>} &
$\m{offer} \; l_i; \, K$ &
Offer the $l_i$ branch \\

&
&
$\m{case} \; a \; \{ \overline{ l_i: K_i } \}$ &
Branch to continuation $K_i$ when channel $a$ offers $l_i$. \\

- &
\texttt{Rec<F>} &
\texttt{fix\_session(cont)} &
Fold session type \texttt{F::Applied} offered by \texttt{cont}. \\

&
&
\texttt{unfix\_session(a, cont)} &
Unfold channel \texttt{a} to session type \texttt{F::Applied} in \texttt{cont}. \\

\bottomrule
\end{tabular}
\end{scriptsize}
\end{table*}

\subsection{Typing Rules for \SILLR}

Following is a list of inference rules in \SILLR.

\noindent\textbf{Communication}

\begin{footnotesize}
\begin{minipage}[c]{0.5\textwidth}
\begin{center}
\[
\inferrule*[right=(\text{\textsc{T-cut}})]
{ \Gamma \, ; \, \Delta_1 \entails a :: A
  \\
  \Gamma \, ; \, \Delta_2, a': A \entails b :: B
}
{ \begin{array}{c}
  \Gamma \, ; \, \Delta_1, \Delta_2 \entails
  a' \, \leftarrow \, \m{cut} \; a \, ; \; b \, :: \, B
  \end{array}
}
\]
\end{center}

\end{minipage}
\begin{minipage}[c]{0.5\textwidth}

\begin{center}
\[
\inferrule*[right=(\text{\textsc{T-incl}})]
{ \Gamma \, ; \, \cdot \entails a :: A
  \\
  \Gamma \, ; \, \Delta, a': A \entails b :: B
}
{ \begin{array}{c}
  \Gamma \, ; \, \Delta \entails
  a' \, \leftarrow \, \m{include} \; a \, ; \; b \, :: \, B
  \end{array}
}
\]
\end{center}
\end{minipage}
\end{footnotesize}

\begin{footnotesize}
\begin{minipage}[c]{0.5\textwidth}
\begin{center}
\[
\inferrule* [right=($\text{\textsc{T-app}}$)]
{ \Gamma \, ; \, \cdot \entails f :: A \chanin B \\
  \Gamma \, ; \, \cdot \entails a :: A
}
{ \Gamma \, ; \, \cdot \entails \,
  \m{apply\_channel} \; f \; a \, :: \, B
}
\]
\end{center}

\end{minipage}
\begin{minipage}[c]{0.5\textwidth}

\begin{center}
\[
\inferrule*[right=(\text{\textsc{T-fwd}})]
{ }
{ \Gamma \, ; a: A \entails
  \m{forward} \, a \, :: \, A
}
\]
\end{center}
\end{minipage}
\end{footnotesize}

\noindent\textbf{Termination}

\begin{footnotesize}
\begin{minipage}[c]{0.5\textwidth}
\begin{center}
\[
\inferrule*[right=(\text{\textsc{T1$_\m{R}$}})]
{ }
{ \Gamma \, ; \cdot \entails
  \m{terminate}; \, :: \, \epsilon
}
\]
\end{center}

\end{minipage}
\begin{minipage}[c]{0.5\textwidth}

\begin{center}
\[
\inferrule*[right=(\text{\textsc{T1$_\m{L}$}})]
{ \Gamma \, ; \, \Delta  \entails K :: A
}
{ \Gamma \, ; \, \Delta, \, a: \epsilon \entails \m{wait} \, a; \, K :: A
}
\]
\end{center}
\end{minipage}
\end{footnotesize}

\noindent\textbf{Receive Value}

\begin{footnotesize}
\begin{minipage}[c]{0.5\textwidth}
\begin{center}
\[
\inferrule*[right=(\textsc{T$\triangleright_\m{R}$})]
{
  \Gamma, \, \mi{x}: \tau \, ; \Delta \entails
  K \, :: \, A
}
{ \Gamma \, ; \Delta \entails
  \mi{x} \leftarrow \m{receive\_value}; \, K \, :: \,
  \tau \recvval A
}
\]
\end{center}

\end{minipage}
\begin{minipage}[c]{0.5\textwidth}

\begin{center}
\[
\inferrule*[right=(\textsc{T$\triangleright_\m{L}$})]
{\Gamma \, ; \, \Delta, a: A \entails K :: B}
{\Gamma, \, x: \tau ; \, \Delta, \, a: \tau \recvval A \entails
  \m{send\_value\_to} \; a \; x ; \, K :: B}
\]
\end{center}
\end{minipage}
\end{footnotesize}

\noindent\textbf{Send Value}

\begin{footnotesize}
\begin{minipage}[c]{0.5\textwidth}
\begin{center}
\[
\inferrule*[right=(\textsc{T$\triangleleft_\m{R}$})]
{\Gamma \, ; \, \Delta \entails K :: A}
{\Gamma, \, x: \tau ; \, \Delta \entails
  \m{send\_value} \; x ; \, K :: \tau \sendval A}
\]
\end{center}

\end{minipage}
\begin{minipage}[c]{0.5\textwidth}
\begin{center}
\[
\inferrule*[right=(\textsc{T$\triangleleft_\m{L}$})]
{ \Gamma, \, \mi{a}: \tau \, ; \Delta, a: A \entails
  K \, :: \, B
}
{ \Gamma \, ; \Delta, a: \tau \recvval A \entails
  \mi{x} \leftarrow \m{receive\_value\_from} \, a; \, K \, :: \, B
}
\]
\end{center}
\end{minipage}
\end{footnotesize}

\noindent\textbf{Receive Channel}

\begin{footnotesize}
\begin{minipage}[c]{0.5\textwidth}
\begin{center}
\[
\inferrule*[right=($\textsc{T$\chanin_\m{R}$}$)]
{\Gamma \, ; \, \Delta, a: A \entails K :: B}
{ \Gamma \, ; \, \Delta \entails
  a \, \leftarrow \, \m{receive\_channel} ; \, K :: A \chanin B
}
\]
\end{center}

\end{minipage}
\begin{minipage}[c]{0.5\textwidth}

\begin{center}
\[
\inferrule*[right=(\textsc{T$\chanin_\m{L}$})]
{\Gamma \, ; \, \Delta, f: A_2 \entails K :: B}
{\Gamma \, ; \, \Delta, \, f: A_1 \chanin A_2, \, a: A_1 \entails
  \m{send\_channel\_to} \; f \; a ; \, K :: B}
\]
\end{center}
\end{minipage}
\end{footnotesize}

\noindent\textbf{Send Channel}

\begin{footnotesize}
\begin{minipage}[c]{0.5\textwidth}
\begin{center}
\[
\inferrule*[right=($\textsc{T$\chanout_\m{R}$}$)]
{\Gamma \, ; \, \Delta \entails K :: B}
{ \Gamma \, ; \, \Delta, a: A \entails
  \m{send\_channel\_from} \; a ; \, K :: A \chanout B
}
\]
\end{center}

\end{minipage}
\begin{minipage}[c]{0.5\textwidth}

\begin{center}
\[
\inferrule*[right=(\textsc{T$\chanout_\m{L}$})]
{\Gamma \, ; \, \Delta, f: A_2, \, a: A_1 \entails K :: B}
{\Gamma \, ; \, \Delta, \, f: A_1 \chanout A_2 \entails
  a \leftarrow \m{receive\_channel\_from} \; f ; \, K :: B}
\]
\end{center}
\end{minipage}
\end{footnotesize}

\noindent\textbf{Shared Session Types}

\begin{footnotesize}
\begin{minipage}[c]{0.5\textwidth}
\begin{center}
\[
\inferrule*[right=($\textsc{T$\lineartoshared{}_\m{R}$}$)]
{\Gamma \, ; \, \cdot \entails K :: A}
{ \Gamma \, ; \, \cdot \entails
  \m{accept\_shared\_session} ; \, K :: \lineartoshared A
}
\]

\[
\inferrule*[right=($\textsc{T$\sharedtolinear{}_\m{R}$}$)]
{\Gamma \, ; \, \cdot \entails K :: S}
{ \Gamma \, ; \, \cdot \entails
  \m{detach\_shared\_session} ; \, K :: \sharedtolinear S
}
\]
\end{center}

\end{minipage}
\begin{minipage}[c]{0.5\textwidth}

\begin{center}
\[
\inferrule*[right=($\textsc{T$\lineartoshared{}_\m{L}$}$)]
{\Gamma \, ; \, \Delta, \, a: A \entails K :: B}
{ \Gamma, s: \lineartoshared A \, ; \, \Delta \entails
  a \leftarrow \m{acquire\_shared\_session} \, s; \, K :: B
}
\]

\[
\inferrule*[right=($\textsc{T$\sharedtolinear{}_\m{L}$}$)]
{\Gamma, s: S \, ; \, \Delta \entails K :: B}
{ \Gamma \, ; \, \Delta, \, a: \sharedtolinear S \entails
  s \leftarrow \m{release\_shared\_session} \, a; \, K :: B
}
\]
\end{center}
\end{minipage}
\end{footnotesize}

\noindent\textbf{External Choice (Binary)}

\begin{footnotesize}
\begin{minipage}[c]{0.5\textwidth}
\begin{center}
\[
\inferrule*[right=($\textsc{T$\extchoice\m{2}_\m{R}$}$)]
{ \Gamma \, ; \, \Delta \entails K_l :: A
  \\
  \Gamma \, ; \, \Delta \entails K_r :: B
}
{ \Gamma \, ; \, \Delta \entails
  \m{offer\_choice} \; K_l \; K_r :: A \extchoice B
}
\]
\end{center}

\end{minipage}
\begin{minipage}[c]{0.5\textwidth}

\begin{center}
\[
\inferrule*[right=(\textsc{T$\extchoice\m{2}_\m{L}$})]
{\Gamma \, ; \, \Delta, a: A_1 \entails K :: B}
{\Gamma \, ; \, \Delta, \, a: A_1 \extchoice A_2 \entails
  \m{choose\_left} \, a ; \, K :: B}
\]

\[
\inferrule*[right=(\textsc{T$\extchoice\m{2}_\m{L}$})]
{\Gamma \, ; \, \Delta, a: A_2 \entails K :: B}
{\Gamma \, ; \, \Delta, \, a: A_2 \extchoice A_2 \entails
  \m{choose\_right} \, a ; \, K :: B}
\]
\end{center}
\end{minipage}
\end{footnotesize}

\noindent\textbf{Internal Choice (Binary)}

\begin{footnotesize}
\begin{minipage}[c]{0.5\textwidth}
\begin{center}
\[
\inferrule*[right=($\textsc{T$\intchoice\m{2}_\m{R}$}$)]
{ \Gamma \, ; \, \Delta \entails K :: A
}
{ \Gamma \, ; \, \Delta \entails
  \m{offer\_left}; \, K :: A \intchoice\m{2} B
}
\]

\[
\inferrule*[right=($\textsc{T$\intchoice\m{2}_\m{R}$}$)]
{ \Gamma \, ; \, \Delta \entails K :: B
}
{ \Gamma \, ; \, \Delta \entails
  \m{offer\_right}; \, K :: A \intchoice B
}
\]
\end{center}

\end{minipage}
\begin{minipage}[c]{0.5\textwidth}

\begin{center}
\[
\inferrule*[right=(\textsc{T$\intchoice\m{2}_\m{L}$})]
{ \Gamma \, ; \, \Delta, a: A_1 \entails K_l :: B
  \\
  \Gamma \, ; \, \Delta, a: A_2 \entails K_r :: B
}
{\Gamma \, ; \, \Delta, \, a: A_1 \intchoice A_2 \entails
  \m{case} \, a \; K_l \; K_r :: B}
\]

\end{center}
\end{minipage}
\end{footnotesize}

\noindent\textbf{External Choice}

\begin{footnotesize}
\begin{minipage}[c]{0.5\textwidth}
\begin{center}
\[
\inferrule*[right=($\textsc{T$\extchoice_\m{R}$}$)]
{ \Gamma \, ; \, \Delta \entails \overline{ K_i :: A_i }
}
{ \Gamma \, ; \, \Delta \entails
  \m{offer\_choice} \; \{ \overline{ l_i: K_i } \} :: \extchoice\{ \overline{ l_i: A_i } \}
}
\]
\end{center}

\end{minipage}
\begin{minipage}[c]{0.5\textwidth}

\begin{center}
\[
\inferrule*[right=(\textsc{T$\extchoice_\m{L}$})]
{\Gamma \, ; \, \Delta, a: A_i \entails K :: B}
{\Gamma \, ; \, \Delta, \, a: \extchoice\{ \overline{ l_i: A_i } \} \entails
  \m{choose} \; a \; l_i ; \, K :: B}
\]
\end{center}
\end{minipage}
\end{footnotesize}

\noindent\textbf{Internal Choice}

\begin{footnotesize}
\begin{minipage}[c]{0.5\textwidth}
\begin{center}
\[
\inferrule*[right=($\textsc{T$\intchoice_\m{R}$}$)]
{ \Gamma \, ; \, \Delta \entails K :: A
}
{ \Gamma \, ; \, \Delta \entails
  \m{offer} \; l_i \, ; \, K :: \intchoice\{ \overline{ l_i: A_i } \}
}
\]
\end{center}

\end{minipage}
\begin{minipage}[c]{0.5\textwidth}

\begin{center}
\[
\inferrule*[right=(\textsc{T$\intchoice\m{2}_\m{L}$})]
{ \overline{ \Gamma \, ; \, \Delta, a: A_i \entails K_i :: B }
}
{\Gamma \, ; \, \Delta, \, a: \intchoice\{ \overline{ l_i: A_i } \} \entails
  \m{case} \, a \; \{ \overline { l_i:  K_i } \} :: B }
\]

\end{center}
\end{minipage}
\end{footnotesize}

\subsection{Typing Constructs in Ferrite}\label{sec:ferrite-constructs}

Following is a list of function signatures of the
term constructors provided in Ferrite.
\linebreak

\subsubsection{Forward}

\begin{lstlisting}[language=Rust,style=nicerust]
fn forward<N, C, A>(_: N) -> PartialSession<C, A>
where
  A: Protocol,
  C: Context,
  N::Target: EmptyContext,
  N: ContextLens<C, A, Empty>
\end{lstlisting}

\subsubsection{Termination}

\begin{lstlisting}[language=Rust,style=nicerust]
pub struct End;
impl Protocol for End { ... }
\end{lstlisting}

\begin{lstlisting}[language=Rust,style=nicerust]
fn terminate<C>() -> PartialSession<C, End>
where
  C : EmptyContext
\end{lstlisting}

\begin{lstlisting}[language=Rust,style=nicerust]
fn wait<N, C, A>(
  _ : N,
  cont : PartialSession<N::Target, A>
) -> PartialSession<C, A>
where
  C : Context,
  A : Protocol,
  N : ContextLens<C, End, Empty>
\end{lstlisting}

\subsubsection{Communication}

\begin{lstlisting}[language=Rust,style=nicerust]
fn cut<X, C, C1, C2, A, B>(
  cont1 : PartialSession<C1, A>,
  cont2 : impl FnOnce(C2::Length) -> PartialSession<C2::Appended, B>
) -> PartialSession<C, B>
where
  A : Protocol,
  B : Protocol,
  C : Context,
  C1 : Context,
  C2 : Context,
  X : SplitContext<C, Left = C1, Right = C2>,
  C2 : AppendContext<(A, ())>

fn include_session<C, A, B>(
  session : Session<A>,
  cont : impl FnOnce(C::Length) -> PartialSession<C::Appended, B>
) -> PartialSession<C, B>
where
  A : Protocol,
  B : Protocol,
  C : Context,
  C : AppendContext<(A, ())>

fn apply_channel<A, B>(
  f : Session<ReceiveChannel<A, B>>,
  a : Session<A>,
) -> Session<B>
where
  A : Protocol,
  B : Protocol
\end{lstlisting}

\subsubsection{Receive Value}

\begin{lstlisting}[language=Rust,style=nicerust]
struct ReceiveValue<T, A> { ... }
impl<T, A> Protocol for ReceiveValue<T, A>
where
  T: Send + 'static,
  A: Protocol
{ ... }
\end{lstlisting}

\begin{lstlisting}[language=Rust,style=nicerust]
fn receive_value<T, C, A>(
  cont : impl FnOnce(T) -> PartialSession<C, A> + Send + 'static
) -> PartialSession<C, ReceiveValue<T, A>>
where
  T : Send + 'static,
  A : Protocol,
  C : Context

fn send_value_to<N, C, A, B, T>(
  _ : N,
  val : T,
  cont : PartialSession<N::Target, A>
) -> PartialSession<C, A>
where
  A : Protocol,
  B : Protocol,
  C : Context,
  T : Send + 'static,
  N : ContextLens<C, ReceiveValue<T, B>, B>
\end{lstlisting}

\subsubsection{Send Value}

\begin{lstlisting}[language=Rust,style=nicerust]
struct SendValue<T, A> { ... }
impl<T, A> Protocol for SendValue<T, A>
where
  T: Send + 'static,
  A: Protocol
{ ... }
\end{lstlisting}

\begin{lstlisting}[language=Rust,style=nicerust]
fn send_value<T, C, A>(
  val : T,
  cont : PartialSession<C, A>
) -> PartialSession<C, SendValue<T, A>>
where
  T : Send + 'static,
  A : Protocol,
  C : Context

fn receive_value_from<N, C, T, A, B>(
  _ : N,
  cont : impl FnOnce(T) -> PartialSession<N::Target, B> + Send + 'static
) -> PartialSession<C, B>
where
  A : Protocol,
  B : Protocol,
  C : Context,
  T : Send + 'static,
  N : ContextLens<C, SendValue<T, A>, A>
\end{lstlisting}

\subsubsection{Receive Channel}

\begin{lstlisting}[language=Rust,style=nicerust]
pub struct ReceiveChannel<A, B> { ... }
impl<A: Protocol, B: Protocol> Protocol for ReceiveChannel<A, B> { ... }
\end{lstlisting}

\begin{lstlisting}[language=Rust,style=nicerust]
fn receive_channel<C, A, B>(
  cont : impl FnOnce(C::Length) -> PartialSession<C::Appended, B>
) -> PartialSession<C, ReceiveChannel<A, B>>
where
  A : Protocol,
  B : Protocol,
  C : Context,
  C : AppendContext<(A, ())>

fn send_channel_to<N1, N2, C, A1, A2, B>(
  _ : N1,
  _ : N2,
  cont : PartialSession<N1::Target, B>
) -> PartialSession<C, B>
where
  C : Context,
  A1 : Protocol,
  A2 : Protocol,
  B : Protocol,
  N2 : ContextLens<C, A1, Empty>,
  N1 : ContextLens<N2::Target, ReceiveChannel<A1, A2>, A2>
\end{lstlisting}

\subsubsection{Send Channel}

\begin{lstlisting}[language=Rust,style=nicerust]
struct SendChannel<A, B> { ... }
impl<A: Protocol, B: Protocol> Protocol for SendChannel<A, B>
\end{lstlisting}

\begin{lstlisting}[language=Rust,style=nicerust]
fn send_channel_from<C, A, B, N>(
  _ : N,
  cont : PartialSession<N::Target, B>
) -> PartialSession<C, SendChannel<A, B>>
where
  A : Protocol,
  B : Protocol,
  C : Context,
  N : ContextLens<C, A, Empty>
\end{lstlisting}

\begin{lstlisting}[language=Rust,style=nicerust]
fn receive_channel_from<C1, C2, A1, A2, B, N>(
  _ : N,
  cont_builder : impl FnOnce(C2::Length) -> PartialSession<C2::Appended, B>
) -> PartialSession<C1, B>
where
  A1 : Protocol,
  A2 : Protocol,
  B : Protocol,
  C1 : Context,
  C2 : AppendContext<(A1, ())>,
  N : ContextLens<C1, SendChannel<A1, A2>, A2, Target = C2>
\end{lstlisting}

\subsubsection{External Choice}

\begin{lstlisting}[language=Rust,style=nicerust]
struct ExternalChoice<Row> { ... }
impl<Row> Protocol for ExternalChoice<Row>
where
  Row: ToRow + Send + 'static
{ ... }
\end{lstlisting}

\begin{lstlisting}[language=Rust,style=nicerust]
fn offer_choice<C, Row1, Row2>(
  cont1: impl for<'r> FnOnce(
      AppSum<'r, Row2, InjectSessionF<'r, Row1, C>>,
    ) -> ContSum<'r, Row1, C>
    + Send
    + 'static
) -> PartialSession<C, ExternalChoice<Row1>>
where
  C: Context,
  Row1: Send + 'static,
  Row2: Send + 'static,
  Row1: ToRow<Row = Row2>,
  Row2: RowCon,
  Row2: SumFunctor
\end{lstlisting}

\begin{lstlisting}[language=Rust,style=nicerust]
fn choose<N, M, C1, C2, A, B, Row1, Row2>(
  _: N,
  _: M,
  cont: PartialSession<C2, A>
) -> PartialSession<C1, A>
where
  C1: Context,
  C2: Context,
  A: Protocol,
  B: Protocol,
  Row2: RowCon,
  Row1: Send + 'static,
  Row2: Send + 'static,
  Row1: ToRow<Row = Row2>,
  N: ContextLens<C1, ExternalChoice<Row1>, B, Target = C2>,
  M: Prism<Row2, Elem = B>
\end{lstlisting}

\subsubsection{Internal Choice}

\begin{lstlisting}[language=Rust,style=nicerust]
struct InternalChoice<Row> { ... }
impl<Row> Protocol for InternalChoice<Row>
where
  Row: ToRow + Send + 'static
{ ... }
\end{lstlisting}

\begin{lstlisting}[language=Rust,style=nicerust]
fn offer_case<N, C, A, Row1, Row2>(
  _: N,
  cont: PartialSession<C, A>
) -> PartialSession<C, InternalChoice<Row1>>
where
  C: Context,
  A: Protocol,
  Row1: Send + 'static,
  Row2: Send + 'static,
  Row2: RowCon,
  Row1: ToRow<Row = Row2>,
  N: Prism<Row2, Elem = A>
\end{lstlisting}

\begin{lstlisting}[language=Rust,style=nicerust]
fn case<N, C1, C2, B, Row1, Row2>(
  _: N,
  cont1: impl for<'r> FnOnce(
      AppSum<'r, Row2, ContF<'r, N, C2, B>>,
    ) -> ChoiceRet<'r, N, C2, B>
    + Send
    + 'static
) -> PartialSession<C1, B>
where
  B: Protocol,
  C1: Context,
  C2: Context,
  Row1: Send + 'static,
  Row2: Send + 'static,
  Row1: ToRow<Row = Row2>,
  N: ContextLens<C1, InternalChoice<Row1>, Empty, Target = C2>
\end{lstlisting}

\subsubsection{Recursive Session Types}

\begin{lstlisting}[language=Rust,style=nicerust]
fn fix_session<R, F, A, C>(
  cont: PartialSession<C, A>
) -> PartialSession<C, RecX<R, F>>
where
  C: Context,
  R: Context,
  F: Protocol,
  A: Protocol,
  F: RecApp<(RecX<R, F>, R), Applied = A>
\end{lstlisting}

\begin{lstlisting}[language=Rust,style=nicerust]
fn unfix_session<N, C1, C2, A, B, R, F>(
  _n: N,
  cont: PartialSession<C2, B>
) -> PartialSession<C1, B>
where
  B: Protocol,
  C1: Context,
  C2: Context,
  F: Protocol,
  R: Context,
  F: RecApp<(RecX<R, F>, R), Applied = A>,
  A: Protocol,
  N: ContextLens<C1, RecX<R, F>, A, Target = C2>
\end{lstlisting}

\subsubsection{Shared Session Types}\label{sec:shared-ferrite-constructs}

\begin{lstlisting}[language=Rust,style=nicerust]
struct LinearToShared<F> { ... }
struct SharedToLinear<F> { ... }
struct Lock<F> { ... }

impl<F> SharedProtocol for LinearToShared<F>
where
  F: Protocol,
  F: SharedRecApp<SharedToLinear<LinearToShared<F>>>,
  F::Applied: Protocol
{ ... }

impl<F> Protocol for SharedToLinear<LinearToShared<F>>
where
  F: SharedRecApp<SharedToLinear<LinearToShared<F>>> + Send + 'static
{ ... }

impl<F> Protocol for Lock<F>
where
  F: Protocol,
  F: SharedRecApp<SharedToLinear<LinearToShared<F>>>,
  F::Applied: Protocol
{ ... }
\end{lstlisting}

A detail we omitted in the main text is that we introduced a special linear session type called
\code{Lock}, internal to the library. The \code{Lock} type holds the
underlying shared Rust channel that connects to the corresponding endpoint held by
\code{SharedChannel}. This allows multiple uses of \code{accept\_shared\_session} and
\code{detach\_shared\_session} to all access the same underlying Rust channel
without having to rely on global state.

An additional role of the linear session type \code{Lock} is that it
also enforces the equi-synchronizing constraint of shared session
type, by requiring all use of \code{accept\_shared\_session} to always
be accompanied by \code{detach\_shared\_session} with the same shared
session type.  This provides the same functionality of enforcing the
equi-synchronizing constraint as specified in Section 3.3
in~\cite{BalzerICFP2017}.

\begin{lstlisting}[language=Rust, style=nicerust]
fn accept_shared_session<F>(
  cont: impl Future<Output = PartialSession<(Lock<F>, ()), F::Applied>>
    + Send
    + 'static
) -> SharedSession<LinearToShared<F>>
where
  F: Protocol,
  F: SharedRecApp<SharedToLinear<LinearToShared<F>>>,
  F::Applied: Protocol
\end{lstlisting}

The \code{accept\_shared\_session} construct is parameterized over a
shared session type \code{LinearToShared<F>}.  The type \code{F} is
required to implement
\code{SharedRecApp<SharedToLinear<LinearToShared<F>>>}, which unfolds
the shared session type by applyig the type
\code{SharedToLinear<LinearToShared<F>>} to \code{F}.  The
continuation is an \code{async} block with \code{PartialSession} result that
offers the linear session type \code{F::Applied}.  It also has an
internal session type \code{Lock<F>}, which is described next.  The
construct returns a shared session type program of type
\code{SharedSession<LinearToShared<F>>}.  This needs to be passed to
\code{run\_shared\_session} to execute the program and get back a
shared channel of type \code{SharedChannel<LinearToShared<F>>}.

\begin{lstlisting}[language=Rust, style=nicerust]
fn detach_shared_session<F, C>(
  cont: SharedSession<LinearToShared<F>>
) -> PartialSession<(Lock<F>, C), SharedToLinear<LinearToShared<F>>>
where
  F: Protocol,
  F: SharedRecApp<SharedToLinear<LinearToShared<F>>>,
  F::Applied: Protocol,
  C: EmptyContext
\end{lstlisting}

The \code{detach\_shared\_session} construct is parameterized by a
linear session type \code{LinearToShared<F>} and an empty linear
context \code{C}.  The type \code{F} is required to implement
\code{SharedRecApp<SharedToLinear<LinearToShared<F>>>} to unfold
\code{F} recursively. This is required for \code{LinearToShared<F>} to
satisfy the \code{SharedProtocol} constraint.  The construct accepts a
\code{SharedSession} continuation with the offered shared session type
\code{LinearToShared<F>}.  Note that this is the only continuation
that is \textit{not} a \code{PartialSession}.  It is also \textit{not}
a \code{SharedChannel}, as this is a shared Ferrite program that is
yet to be executed.  The construct returns a \code{PartialSession}
that offers the linear session type \code{SharedToLinear<F>}. It also
has a linear context with \code{Lock<F>} being the first linear
channel, and the tail \code{C} being an empty linear context of
arbitrary length.

\begin{lstlisting}[language=Rust, style=nicerust]
fn acquire_shared_session<C, F, A>(
  shared: SharedChannel<LinearToShared<F>>,
  cont1: impl FnOnce(C::Length) -> PartialSession<C::Appended, A> + Send + 'static
) -> PartialSession<C, A>
where
  C: Context,
  F: Protocol,
  A: Protocol,
  F::Applied: Protocol,
  F: SharedRecApp<SharedToLinear<LinearToShared<F>>>,
  C: AppendContext<(F::Applied, ())>
\end{lstlisting}

The \code{acquire\_shared\_session} construct is parameterized over a
shared session type \linebreak \code{LinearToShared<F>}, a linear
context \code{C}, and an offered session type \code{A}.  The type
\code{F} is required to implement
\code{SharedRecApp<SharedToLinear<LinearToShared<F>>>}, which unfolds
the shared session type by applying the type
\code{SharedToLinear<LinearToShared<F>>} to \code{F}. The unfolded
session type \code{F::Applied} is a linear session type implementing
\code{Protocol}, and it is appended to the end of \code{C} using
\code{AppendContext}, with \code{C::Appended} being the result.

The first argument to \code{acquire\_shared\_session} is a cloneable
\code{SharedChannel} of  (shared) session type
\code{LinearToShared<F>}. The second argument is the continuation
closure.  It is given the context lens \code{C::Length}, which
implements the context lens to access the linear channel
\code{F::Applied} in \code{C::Appended}.  The continuation closure
returns a \code{PartialSession} with \code{C::Appended} being the
linear context, and \code{A} being the offered session type.  The
construct returns a \code{PartialSession} that works with the original
linear context \code{C}, and offers the session type \code{A}.

\begin{lstlisting}[language=Rust, style=nicerust]
fn release_shared_session<N, C1, C2, A, B>(
  _n: N,
  cont: PartialSession<C2, B>,
) -> PartialSession<C1, B>
where
  A: Protocol,
  B: Protocol,
  C1: Context,
  C2: Context,
  A: SharedRecApp<SharedToLinear<LinearToShared<A>>>,
  N: ContextLens<C1, SharedToLinear<LinearToShared<A>>, Empty, Target = C2>
\end{lstlisting}

The \code{release\_shared\_session} construct is parameterized over a
linear session type \code{SharedToLinear<A>}, a linear context
\code{C}, a context lens \code{N} for accessing
\code{SharedToLinear<LinearToShared<A>>} from \code{C}, and an offered
session type \code{B}. The continuation is a \code{PartialSession}
with \code{N::Target} being the linear context \code{C} with
\code{SharedToLinear<LinearToShared<A>>} removed, and offers the
session type \code{B}.  The construct returns a \code{PartialSession}
with the original linear context \code{C}, and offers the session type
\code{B}.


\section{Dynamics}\label{sec:dynamics}

\Cref{sec:statics} introduced the type system of Ferrite, based on the constructs \code{End},
\code{ReceiveValue}, and \code{ReceiveChannel}.  This section revisits those constructs and fills in
the missing implementations to make the constructs executable, amounting to the \textit{dynamic semantics} of Ferrite.

\subsection{One-shot Channels}

Internally, Ferrite uses \code{tokio}'s \code{oneshot} \cite{TokioWebsite} channels
as the primitive building block for session-typed channels. A one-shot channel with
a payload type \code{P} is consist of a pair of sender and receiver, of type
\code{Sender<P>} and \code{Receiver<P>}, resp., denoting the two endpoints
of the channel. The channel is one-shot in the sense that at most one value of type
\code{P} can be sent across the channel. However since the one-shot channel is affine,
it is also possible to have no value being sent over the channel.

The one-shot channel can be used directly by Rust programmers to emulate
simple session types.
As an example, the session type \code{ReceiveValue<i32, End>} can be implemented using
one-shot channels as follows:

\begin{lstlisting}[language=Rust, style=nicerust]
use tokio::{task, try_join};
use tokio::sync::oneshot::{channel, Sender, Receiver};

async fn receive_int_provider(value_receiver: Receiver<(i32, Sender<()>)>) {
  let (value, end_sender) = value_receiver.await.unwrap()
  println!("provider received value: {}", value);
  end_sender.send(()).unwrap();
}
async fn receive_int_client(value_sender: Sender<(i32, Receiver<()>)>) {
  let (end_sender, end_receiver) = channel();
  value_sender.send((42, end_sender));
  end_receiver.await.unwrap();
}
async fn main() {
  let (value_sender, value_receiver) = channel();
  let child1 = spawn(async move {
    receive_int_provider(value_receiver).await;
  });
  let child2 = spawn(async move {
    receive_int_client(value_sender).await;
  });
}
\end{lstlisting}

The code above defines the \code{receive\_int\_provider} and
\code{receive\_int\_client} functions to execute the provider and
client processes corresponding to the session type
\code{ReceiveValue<i32, End>}, resp. On the provider side, it
needs to first receive an \code{i32} value and then send back an end
signal to the client when it is terminating. This corresponds to the
one-shot channel type \code{Receiver<(i32, Sender<()>)>}, with the
\code{Sender<()>} used to send a unit \code{()} as termination
signal. On the receiver side, the polarity of the one-shot channel is
switched and become \code{Sender<(i32, Receiver<()>)>}. This indicates
that the client first sends an \code{i32} value, together with a
\code{Receiver<()>} for the provider to send back the termination
signal.

\subsection{Protocol Definitions}

The above example demonstrates that even a simple session type like
\code{ReceiveValue<i32, End>} requires non-trivial effort to be
implemented manually using one-shot channels. To automate this in
Ferrite, we need to derive the one-shot channel types
\code{Receiver<(i32, Sender<()>)>} and \code{Sender<(i32,
  Receiver<()>)>} from the session type \code{ReceiveValue<i32, End>}.
This is achieved by defining some associated types and methods in the
\code{Protocol} trait:

\begin{lstlisting}[language=Rust, style=nicerust]
trait Protocol {
  type ProviderEndpoint;
  type ClientEndpoint;

  fn create_endpoints() -> (Self::ProviderEndpoint, Self::ClientEndpoint);
}
\end{lstlisting}

The associated types \code{ProviderEndpoint} and \code{ClientEndpoint}
are used to define the one-shot channel types for the provider end and
consumer end, resp.  The trait method \code{create\_endpoints}
is used to create a channel pair which connects both the provider and
client endpoints.  Following the previous example, the implementation
should derive the type \code{<ReceiveValue<i32,
  End>>::ProviderEndpoint} to be \code{Receiver<(i32, Sender<()>)>},
and \code{<ReceiveValue<i32, End>>::ClientEndpoint} to be
\code{Sender<(i32, Receiver<()>)>}. This is implemented by first
implementing \code{Protocol} for \code{End}:

\begin{lstlisting}[language=Rust, style=nicerust]
impl Protocol for End
{
  type ProviderEndpoint = Sender<()>;
  type ClientEndpoint = Receiver<()>;

  fn create_endpoints() -> (Self::ProviderEndpoint, Self::ClientEndpoint)
  {
    channel()
  }
}
\end{lstlisting}

In the implementation of the \code{End} protocol, the provider end is the party
that needs to send the termination signal \code{()} to the client end. Hence
its \code{ProviderEndpoint} type is \code{Sender<()>}, and vice versa for the
client end. The implementation of the \code{create\_endpoints} method is to
simply call \code{channel()} to create the one-shot channel pair.

To implement \code{Protocol} for a session type \code{ReceiveValue<T, A>}, we would
need to make use of the \code{Protocol} implementation for the continuation
session type \code{A}:

\begin{lstlisting}[language=Rust, style=nicerust]
impl<T, A> Protocol for ReceiveValue<T, A>
{
  type ProviderEndpoint = Receiver<(T, A::ProviderEndpoint)>;
  type ClientEndpoint = Sender<(T, A::ProviderEndpoint)>;

  fn create_endpoints() -> (Self::ProviderEndpoint, Self::ClientEndpoint)
  {
    let (sender, receiver) = channel();
    (receiver, sender)
  }
}
\end{lstlisting}

The provider end is given a receiver for the value \code{T}, together
with its continuation endpoint for \code{A}. Given that the
continuation for the provider also needs the provider endpoint, and it
has to be extracted from the receiver, the provider would need to
receive \code{A::ProviderEndpoint} alongside with the value
\code{T}. Hence the associated type \code{<ReceiveValue<T,
  A>>::ProviderEndpoint} becomes \code{Receiver<(T,
  A::ProviderEndpoint)>}. On the client side, the value \code{T} needs
to be sent alongside with \code{A::ProviderEndpoint}, hence the
associated type \code{<ReceiveValue<T, A>>::ClientEndpoint} is
\code{Sender<(T, A::ProviderEndpoint)>}. Notice that both the
\code{ProviderEndpoint} and \code{ClientEndpoint} associated types for
\code{ReceiveValue<T, A>} contains \code{A::ProviderEndpoint}, but not
\code{A::ClientEndpoint}.

In the implementation of \code{create\_endpoints} for
\code{ReceiveValue}, the ordering of the sender and receiver pair
returned from calling \code{channel()} is flipped. This is because
\code{create\_endpoints} always return the provider endpoint first
followed by the client endpoint. And since the provider endpoint is a
receiver in this case, it needs to be returned in the first position.

With the \code{Protocol} definitions of both \code{End} and
\code{ReceiveValue}, we can follow that the associated types and
channel creation for \code{ReceiveValue<i32, End>} matches the channel
types and behavior of the example at the beginning of this section.

\subsection{Linear Context}

The linear context of a Ferrite program comprises the client endpoints
for the session types. Conceptually, Ferrite needs to derive from a
session type list \code{HList![A0, A1, ...]} into a list of client
endpoint list \code{HList![A0::ClientEndpoint, A1::ClientEndpoint,
  ...]}. However a linear context may also contain the special
\code{Empty} element, which do not implement \code{Protocol}. To allow
the transformation of the linear context, we need to first add an
associated type to the \code{Slot} trait as follows:

\begin{lstlisting}[language=Rust, style=nicerust]
trait Slot {
  type Endpoint: Send;
}
impl<A: Protocol> Slot for A {
  type Endpoint = A::ClientEndpoint;
}
impl Slot for Empty {
  type Endpoint = ();
}
\end{lstlisting}

We define the associated type \code{Endpoint} in \code{Slot} such that
if a type \code{A} implements \code{Protocol}, then \code{A::Endpoint}
is simply \code{A::ClientEndpoint}. We also define the special case
for \code{Empty}, which the \code{Endpoint} associated type is
\code{()} to represent the absence of a client endpoint. With that, we
can extend the \code{Context} trait to include the \code{Endpoints}
associated type:

\begin{lstlisting}[language=Rust, style=nicerust]
trait Context {
  type Endpoints;
}
impl Context for () {
  type Endpoints = ();
}
impl<A: Slot, C: Context> Context for (A, C) {
  type Endpoints = (A::Endpoint, C::Endpoints);
}
\end{lstlisting}

For the base case of an empty list \code{()} (\code{HList![]}), the
result \code{Endpoints} is also an empty list. For the inductive case,
if the tail \code{C} of a linear context \code(A, C) implements
\code{Context}, and the head \code{A} implements \code{Slot}, then the
associated type \code{(A, C)::Endpoints} is \code{(A::Endpoint,
  C::Endpoints)}.

\subsection{Session Dynamics}

Ferrite generates session type programs by composing
\code{PartialSession} objects generated by constructs such as
\code{receive\_value}. To enable execution of the Ferrite program, the
\code{PartialSession} struct contains an internal \code{executor}
field that is defined as follows:

\begin{lstlisting}[language=Rust, style=nicerust]
struct PartialSession<C: Context, A: Protocol> {
  executor: Box<
    dyn FnOnce(
        C::Endpoints,
        A::ProviderEndpoint,
      ) -> Pin<Box<dyn Future<Output = ()> + Send>>
      + Send,
  >
}
\end{lstlisting}

The \code{executor} field contains an \code{FnOnce} closure that
accepts two arguments -- the endpoints for the linear context
\code{C::Endpoints}, and the provider endpoint for the
offered session type \code{A::ProviderEndpoint}.
When called, the closure executes asynchronously by returning a
\textit{future} with the type \code{Pin<Box<dyn Future<Output = ()> + Send>>}.
The boilerplate signature is required, as Rust has not stabilized
the syntactic sugar for async closures. Conceptually,
the closure signature is equivalent to the async function signature
\code{async fn(C::Endpoints, A::ProviderEndpoint)}.

Ferrite keeps the \code{executor} field private within the
library to prevent end users from constructing new
\code{PartialSession} values or running the \code{executor}
closure. This is because the creation and execution
of \code{PartialSession} may be unsafe. We demonstrate two simple
examples of unsafe (\ie non-linear) usage of \code{PartialSession}.

Below shows an example Ferrite program \code{p1} of type
\code{Session<SendValue<String, End>>} is constructed,
but in the \code{executor} closure both the
client endpoints and the provider endpoint are ignored. As a result,
\code{p1} violates the linearity constraint of session
types and never sends any string value or signal for termination.

\begin{lstlisting}[language=Rust, style=nicerust]
let p1: Session<SendValue<String, End>>
  = PartialSession { executor: Box::pin(async |_ctx, _provider_end| { }) };
\end{lstlisting}

Below shows an example client, which calls
a Ferrite program \code{p2} of type
\code{ReceiveValue<String, End>} by directly running its
\code{executor}. The client creates an endpoint pair
but drops the client endpoint. It then executes \code{p2} with the
provider endpoint. However because the client endpoint is dropped,
\code{p2} fails to receive any value, and the program results in a deadlock.

\begin{lstlisting}[language=Rust, style=nicerust]
let p2: Session<ReceiveValue<String, End>> = ...;
let (provider_end, _client_end) = <ReceiveValue<String, End>>::create_endpoints();
(p2.executor)((), provider_end).await;
\end{lstlisting}

From the examples above we can see that direct access to the
\code{executor} field is unsafe. The \code{PartialSession}
is used with care within Ferrite to ensure that linearity
is enforced in the implementation. Externally, the
\code{run\_session} function is provided for executing
Ferrite programs of type \code{Session<End>}, as
only such programs can be executed safely without
additional safe guard.


\section{Rust as a Host Language}

In this section, we address some common questions arise from the
choice of using Rust as a host language for Ferrite.

\subsection{Benefits of Affine Type System}

The affine type system in Rust helps Ferrite to better verify the
correctness of its underlying implementation. Internally, Ferrite uses
one-shot Rust channels to implement the communication. The affine
property in Rust helps us guarantee that our underlying implementation
cannot accidentally send two payloads through the one-shot channels.

Ferrite user programs also benefit from the affine type system in
Rust.  Ferrite constructs accept continuation closures with the
\code{FnOnce} trait bound, to guarantee that the continuation cannot
be called more than once.  As a result, Rust values can be moved
inside the continuation closures and work more efficiently without
requiring copies to be made. Similarly, the send/receive value
constructs works with the affine type system in Rust, so values such
as byte arrays can be sent efficiently in Ferrite without requiring
copying.

In comparison, while previous works in Haskell and OCaml are able to
enforce the linear usage in session type programs, the structural
semantics of these languages may impose challenge on the compiler from
being able to optimize the use of linear resources inside the
program. In particular, the indexed monad that encapsulates the
session type program is itself copyable.  As a result, continuations
cannot guarantee that the variables they capture cannot be used more
than once.

\subsection{Support for Lifetime}

At the moment, Ferrite requires the continuations to have
\code{'static} lifetime.  This is due to the underlying async
implementations requiring spawned async tasks to have \code{'static}
lifetime. We plan to overcome this limitation in the future by finding
ways to spawn async tasks with a scoped lifetime. Once that limitation
is overcome, it will also be possible to access mutable references
inside scoped Ferrite programs.

\subsection{Type Errors}

Type error messages in Ferrite are expressed in terms of the structs
and traits of Ferrite.  As a result it is not difficult for users to
read and understand the error messages, provided they are familiar
with the basic terminology used by Ferrite.

Consider the example \code{hello\_client} from section \ref{sec:statics}

\begin{lstlisting}[language=Rust, style=nicerust]
let hello_client: Session<
  ReceiveChannel<ReceiveValue<String, End>, End>>
  = receive_channel(| a | {
      send_value_to(a, "Alice".to_string(),
        wait(a, terminate())
      ) });
\end{lstlisting}

If we were to forget to wait for channel \code{a} and terminate immediately,
the following error is generated:

\begin{lstlisting}[language=Rust, style=nicerust]
let hello_client: Session<
  ReceiveChannel<ReceiveValue<String, End>, End>>
  = receive_channel(| a | {
      send_value_to(a, "Alice".to_string(),
        // the trait `EmptyContext` is not implemented for `(End, ())`
        terminate()
      ) });
\end{lstlisting}

This indicates that the linear context \code{(End, ())} is not empty, and as a result the
\code{terminate} construct cannot be used.

If we try to wait for \code{a} to terminate before sending a value to \code{a}, we get a
different error:

\begin{lstlisting}[language=Rust, style=nicerust]
let hello_client: Session<
  ReceiveChannel<ReceiveValue<String, End>, End>>
  = receive_channel(| a | {
      // the trait `ContextLens<(ReceiveValue<String, End>, ()), End, Empty>`
      // is not implemented for `Z`
      wait(a, terminate())
    });
\end{lstlisting}

The error message indicates an invalid use of a context lens to update a channel of the
wrong session type in the linear context. Recall from section 3.2 that the constraint
\code{Z: ContextLens<(ReceiveValue<String, End>, ()), End, Empty>} would require the
first channel (\code{Z}) in the linear context (\code{(ReceiveValue<String, End>, ())})
to be of session type \code{End}, but here the session type of the first channel
in the linear context is \code{ReceiveValue<String, End>}.

Error messages such as the above are commonly generated by non-linear use
of channels or a mismatch in session types. While they require some
understanding of the concepts such as linear context and context lenses,
the error messages are not too difficult to decipher.

\subsection{Hole Driven Development}

Aside from designing readable error messages, we recommend a
\textit{hole-driven} approach of writing Ferrite programs to
minimize the chance of users encountering complex type errors.
In this approach, the user would implement a Ferrite program
in small steps, with the continuation filled with \code{todo!()}
as a placeholder. We demonstrate this by showing how a
new user would implement the \code{hello\_provider} program
in section \ref{sec:statics}:

\begin{lstlisting}[language=Rust, style=nicerust]
let hello_provider: Session<SendValue<String, End>> = todo!();
\end{lstlisting}

The \code{todo!()} macro allows
us to put a placeholder in unfinished Rust code so that we can try and
compile the code and see if there is any type error. By writing our code
step by step and filling the blank with \code{todo!()}, we can narrow down
the potential places where our code is incorrect.
At this stage, we should be able to compile our program with no error.
This shows that the protocol that we have defined, \code{SendValue<String, End>},
is a valid session type. If we have gotten a compile error otherwise,
it could have been caused by us trying to write an invalid protocol
like \code{SendValue<String, String>}.

We can try to compile our code again, and Rust will accept the code
we have written. However the use of \code{todo!()} does not tell us
how we should continue our program. In Rust,
we could use the unit type \code{()} to deliberately cause
a compile error:

\begin{lstlisting}[language=Rust, style=nicerust]
let hello_provider: Session<SendValue<String, End>> =
  send_value("Hello World!".to_string(), ());
\end{lstlisting}

Now if we compile our code, we would get a compile error from Rust:

\begin{lstlisting}[language=Rust, style=nicerust]
error[E0308]: mismatched types
 |
 |  send_value("Hello World!".to_string(), ());
 |                                         ^^ expected struct PartialSession, found ()
 |
 = note: expected struct PartialSession<(), End>
         found unit type ()
\end{lstlisting}

With this compile error, we can know that we are supposed to fill in the hole
with Rust expression that has the type \code{PartialSession<(), End>}.
Sometimes we may also intuitively think of a type that should be in a hole.
In such case, we can also use the \code{todo!() as T} pattern to verify if our intuition
is correct. So we can for example write:

\begin{lstlisting}[language=Rust, style=nicerust]
let hello_provider: Session<SendValue<String, End>> =
  send_value("Hello World!".to_string(), todo!() as Session<End>);
\end{lstlisting}

And our code will compile successfully. If we were to annotate it with an
invalid type, such as \code{todo()! as Session<ReceiveValue<String, End>>} again,
Rust will also return a compile error.
Now that we know the continuation needs to have the type \code{Session<End>}, we
can then fill in the blank with \code{terminate()} and complete our program.


\section{Challenges in Using Ferrite on Servo}\label{sec:servo-challenges}

We report on some of the challenges that we faced when implementing the
Servo canvas component in Ferrite in Section \ref{sec:evaluation},
and how the challenges are addressed.

\subsection{Interprocess Communication}

As a browser rendering engine, Servo puts much emphasis on security,
using sandboxing to ensure that malicious web applications cannot
easily compromise a user's computer. A main design outcome of this
emphasis is that the provider and client are executed in separate OS
processes. Regular Rust channels cannot be used for communication
between different processes, because the underlying implementation
requires a common address space.  As a result, Servo uses the
\code{ipc\_channel} crate to create inter-process communication (IPC)
channels for communication between the provider and client of its
components.  The IPC channels in Servo create a local file socket and
serialize the Rust messages to send them over the socket as raw
bytes. This requires the channel payload types to implement the
\code{Serialize} and \code{Deserialize} traits for them to be usable
in the IPC channels.  IPC channels are themselves serializable, so it
is possible to send an IPC channel over another IPC channel.

Since Ferrite internally makes use of \code{tokio} channels for
communication, this presents challenges since they cannot be
serialized and sent through Servo's IPC channels. For the purpose of
the evaluation, we implemented our own serialization of
\code{SharedChannel}.  Our serialization involves creating a
bidirectional pair of opaque (untyped) IPC channels, and forwards all
communication from the regular Rust channels to the IPC channels.
This approach works, albeit inefficiently, as there needs to be two
background tasks in the provider and client processes to perform the
actual serialization and forwarding. We benchmarked the performance of
our implementation, revealing a decrease of about a factor of ten.  We
have not spent much effort on fine-tuning our serialization
implementation because the primary purpose of this study is to show
that the message protocols underlying Servo's canvas component can be
made explicit and verified in Ferrite.

\subsection{Latency in Acquire-Release}

Servo's canvas component has very high performance demands, requiring
the sending of thousands of messages in a few milliseconds.  In our
initial implementation, we found the Ferrite implementation to be
lacking in performance, despite not saturating the CPU usage. A closer
inspection revealed that the bottleneck was in the latency caused by
the acquire-release cycle introduced in the implementation of shared
session types.  In Ferrite, the client of a shared channel needs to
first send an acquire to the shared provider and then wait for the
acknowledgment before it can start communicating through the acquired
linear channel. This round trip latency becomes significant if the
communication frequency is high. Consider two canvas messages being
sent right after each other. In the original design, the second
message can be sent immediately after the first message has been
sent. In the Ferrite implementation, on the other hand, the two
messages are sent in two separate acquire-release cycles,
interspersing additional acquire and release messages and possibly
delays because of blocking acquires.

The latency is aggravated by the use of
IPC channels. Since IPC channels are mapped to file sockets, efficient parallel
communications must be multiplexed among a small number of channels. For the
case of Ferrite shared channels, the multiplexing currently is done by queuing and
forwarding the requests in serial, which can be inefficient.
As a workaround, we batch messages on the client side, such that
trivial messages like \code{LineTo} are stored in a local \code{Vec<CanvasMessage>}
buffer before being sent to the provider in a new \code{Messages} branch in \code{CanvasOps}.
The buffered messages are sent in batch every ten milliseconds, or
when a non-trivial protocol such as \code{GetImageData} is called.
With batching, we have gained enough performance to render
complex canvases smoothly.



\end{document}